\documentclass[letterpaper,prc,preprint]{revtex4}
\usepackage[dvips]{graphicx}
\usepackage{amssymb, latexsym, amsmath, amsfonts, fullpage}
\usepackage{hyperref}
\usepackage{tabularx}
\usepackage{url}

\begin{document}

\title{Nuclear Density Functional Theory and the Equation of State}

\author{Yeunhwan Lim}
\email{ylim@mail.astro.sunysb.edu}
\address{Department of Physics and Astronomy \\
The State University of New York at Stony Brook, Stony Brook, NY, 11790}
%\author{James. M. Lattimer}
%\email{lattimer@mail.astro.sunysb.edu}
%\address{Department of Physics and Astronomy \\
%The State University of New York at Stony Brook, NY, 11790}

\begin{abstract}
A nuclear density functional can be used to find the binding energy and 
shell structure of nuclei and the energy gap in superconducting nuclear 
matter. In this paper, we study the possible application of a nuclear density 
functional theory to nuclear astrophysics. From energy density functional theory, 
we can deduce the interaction between nucleons
to find a rough estimate of the charge radius 
of the specific nuclei. Compared to the Finite-Range Thomas Fermi model, 
we include three-body forces, which might be important at densities several 
times that of nuclear matter density.
We also add the momentum dependent interaction to take 
into account the effective mass of the nucleons. We study matter in the neutron star
crust using the Wigner-Seitz cell method. By constructing the mass-radius
relation of neutron stars and investigating lepton-rich nuclear matter
 in proto-neutron stars, we find that the density 
functional can be used to construct an equation of state of hot dense matter.
\end{abstract}
 
\maketitle
\newpage

\section{Introduction}
Understanding the equation of state (EOS) of dense matter is important to understand 
heavy-ion collision, supernova explosions, and neutron stars.
Neutron stars are believed to be composed of an outer crust, an inner
 crust, an outer core, and an inner core. The outer crust 
($10^{4}\leq\rho\leq 10^{11}\,\text{g cm}^{-3} $) has a lattice of 
neutron-rich nuclei in a gas of free electrons. In the inner crust 
($10^{11}\leq\rho \leq 10^{14}\,\text{g cm}^{-3}$), there are neutron-rich 
nuclei in a gas of free electrons and neutrons. The density of the outer
 core extends to $\sim 5\times 10^{14}\, \text{g cm}^{-3}$ and is a 
homogeneous liquid mainly composed of neutrons, electrons, protons 
and muons. Not much is not known about the inner core ($5\sim 10 \rho_{0}$, 
 where the saturation density is $\rho_{0}=0.16$fm$^{-3}$),
 however, we believe that there are hyperons in the hadronic phase and 
deconfined quark matter\cite{jm2005}. A proper EOS should be able to
 explain nuclear properties in all density ranges.
We can get basic information about nuclear matter from semiempirical 
mass formulas. One such fomula, known as the Liquid Drop
 Model (LDM), determines the binding energy($B$) of nuclei in terms of a 
nuclear matter contribution and various corrections for finite 
nuclei\cite{gmbook}:
\begin{equation}
B(A)=a_{\text{vol}}A + a_{\text{surf}}A^{2/3} + a_{\text{coul}}Z^2 A^{-1/3}
    + a_{\text{sym}}\frac{(N-Z)^2}{A}\,,
\end{equation}
where $a_{\text{vol}}\approx -16$MeV, $a_{\text{surf}}\approx 20$MeV, 
$a_{\text{coul}} \approx 0.751$MeV, $a_{\text{sym}} \approx 
21.4$MeV, $N$ is the number of neutrons, $Z$ is the number of protons, and
A is the total number of nucleons in a nucleus\cite{gmbook}.
 First is the volume term, which is the binding energy of infinite 
nuclear matter. The second term indicates the reduction of binding due to the
 nucleons on the surface. The third term represents the Coulomb energy, which is 
assumed to be that of uniformly-distributed charge. The last term is the 
symmetry energy, representing the decrease in binding energy
for unequal numbers of protons and neutrons. \\
Baym, Bethe, and Pethick (BBP)\cite{ba71} made an EOS in the range of
densities from $4.3\times 10^{11}$g/cm$^{3}$, where neutrons begin to 
drip out of the neuclei,
 to $5\times10^{14}$g/cm$^{3}$ ($2\rho_{0}$) using a 
compressible
 LDM designed to take into account three important features: (i) the free
  neutron
 gas due to neutron drip; (ii) the nuclear surface energy reduced by the 
neutron gas; and (iii) the effect of the nuclear lattice Coulomb energy. 
Baym, Pethick,
 and Sutherland\cite{ba712} extended the previous model in the regime from 
$10^4$g/cm$^{3}$ to neutron drip and applied it to neutron stars and white dwarfs.\\
Oyamatsu\cite{oy1993} studied nuclear shapes and lattice types (spherical,
 cylindrical, slab, cylindricl hole and spherical hole nuclei) at T = 0 MeV with
 parametrized neutron and proton distributions, and performed the Thomas-Fermi
 calculations with four different energy-density functionals. 
Negele and Vautherin\cite{ne1973} used a different method (Hartree-Fock
 approximation) to study the equation of state at T = 0 MeV. They were able to get a 
density-dependent
 Hamiltonian from two-body interactions and found results similar to BBP.\\
There were several efforts to make thermodynamic tables that can be used for
 supernova simulations. One of them is the Lattimer-Swesty EOS (LS)\cite{ls}. Their
 equation of state is an extension of the previous work by Lattimer et al.(LLPR)
\cite{llpr}. The LLPR EOS is based on the LDM. They included the effect of temperature on
 nuclei, the increase of surface energy as temperature increases, the effect of external
 nucleons, and the effect of nuclear excited states. The LS EOS also took into
 account nuclear deformation and the phase transitions from nuclei to uniform nuclear
 matter at subnuclear densities. Another table was given by Shen et al.\cite{sh2003},
 in which they used the field-theoretical model with the Thomas-Fermi method. They
 constructed an EOS of nuclear matter in a wide range of the baryon mass density
 ($\rho_{B}=1.25\times10^{5} - 2.5\times 10^{15}$ g/cm$^{3}$), temperature
 ($T=0 - 100$ MeV), and proton fraction ($Yp= 0 - 0.56$).\\
Recently, Shen et al.\cite{sht20101} used a density-dependent relativistic mean-field
 theory to construct a nuclear EOS. For high- and intermediate-density nuclear
 matter, they employed relativisitic mean field calculations and used the virial
 expansion to study low-density nuclear matter. The table has the range of density
 from $\rho_{B}=10^{-8} - 1.6 \text{fm}^{-3} $, proton fraction 
$\text{Y}_{p} = 0 - 0.56$, and temperature $T=0.16 - 15.8 \text{MeV}$ for high
 and low density nuclear matter.\\
In this paper we use a simple density functional model to decribe both high and low
 nuclear density. To account for the short range of nuclear forces, we use 
a Gaussian form for the interaction. To find the minimum energy of a
nuclear system, we use the Lagrange multiplier method; that is, the chemical
potentials of neutrons and protons are constant in the cells of nuclear systems.
Using this method, we can find the properties of single nuclei and heavy
nuclei with a neutron gas in the neutron star crust. The MIT bag model is used to
see the phase transition at high baryon density ($\rho_{B}>\rho_{0}$).
 The mass-radius relation and the moment of inertia of cold neutron stars are
calculated using the nuclear density functional. Proto-neutron star matter 
with neutrinos is also investigated for a given entropy per baryon. 

\section{Nuclear Density Functional Theory}
\subsection{Energy density functional}
The energy of the nuclear matter can be given by
\begin{equation}
E=T_{kin} + E_{FR} + E_{ZR} + E_{C} + E_{S,L}
\end{equation}
where $T_{kin}$, $E_{FR}$, $E_{ZR}$, $ E_{C}$, and $E_{S,L}$ are the kinetic
 energy, nuclear finite-range interaction, zero-range interaction,
 Coulomb interaction, and spin-orbit coupling respectively.
The kinetic energy contribution from nucleons is simply obtained by
\begin{equation}\label{eq:kin}
T_{kin}=\int d^{3}r\sum_{t}\frac{\hbar^{2}}{2m}\tau_{t} \,,
\end{equation}
where $t$ is the type of nucleons.\\
In this Gaussian nuclear density functional (GNDF) theory, the number and kinetic
 densities are
\begin{equation}
\rho_{t} =\frac{1}{4\pi^{3}\hbar^{3}}\int_{0}^{\infty} 
f_{t} \,d^{3}p \, ; \quad \tau_{t}=\frac{1}{4\pi^{3}\hbar^{5}}\int_{0}^{\infty}
 f_{t}p^{2} \, d^{3}p\,,
\end{equation}
where $f_{t}$ is the Fermi-Dirac density function,
\begin{equation}
f_{t}=\frac{1}{1 + e^{(\epsilon_{t}-\mu_{t})/T}}\,.
\end{equation}
For the finite-range term, we use a Gaussian phenomenological 
model for the nuclear potential,
\begin{align}\label{eq:fr}
E_{FR} & = \sum_{t}\frac{1}{\pi^{3/2}r_{0}^{3}}\int d^{3}r_{1}d^{3}r_{2}
 e^{-r_{12}^{2}/r_{0}^{2}}\,\Big[ V_{1L}\rho_{t}(\vec{r}_{1}) \rho_{t}(\vec{r}_{2})
 + V_{1U}\rho_{t}(\vec{r}_{1}) \rho_{t^{\prime}}(\vec{r}_{2}) \Big] \notag \\
         &+\sum_{t}\frac{1}{\pi^{3/2}r_{0}^{3}}\int d^{3}r_{1}d^{3}r_{2}
 e^{-r_{12}^{2}/r_{0}^{2}}\,\Big[ V_{2L}\rho_{t}^{1+\epsilon}(\vec{r}_{1})
 \rho_{t}^{1+\epsilon}(\vec{r}_{2}) + V_{2U}\rho_{t}^{1+\epsilon}(\vec{r}_{1})
 \rho_{t^{\prime}}^{1+\epsilon}(\vec{r}_{2}) \Big] \\
         &+ \sum_{t}\frac{1}{\pi^{3/2}r_{0}^{3}}\int d^{3}r_{1}d^{3}r_{2}
 e^{-r_{12}^{2}/r_{0}^{2}}\, \biggl[\int d^{3}p_{t1}d^{3}p_{t2} 
f_{t1}f_{t2}V_{3L} p_{12}^{2} + \int d^{3}p_{t1}d^{3}p_{t^{\prime}2}
 f_{t1}f_{t^{\prime}2}V_{3U} p_{12}^{2} \biggr] \notag \,,
\end{align}
where $p_{12}=\vert \vec{p}_{1} -\vec{p}_{2} \vert$,  
$r_{12}=\vert \vec{r}_{1} -\vec{r}_{2} \vert$, $r_{0}$ is the length of 
interaction, and $V_{1L}$, $V_{1U}$, $V_{2L}$, $V_{2U}$, $V_{3L}$, and $V_{3U}$ 
are interaction parameters to be determined. The last term is added to
 explain the effective mass of nucleons in dense matter.\\
The zero-range term in the nuclear force can be regarded as the energy
 contribution from three-body nuclear forces. The three-body force is quite
 important if the baryon density increases beyond two or three times the
 saturation density. One possible form of the three-body force is\cite{gmbook}
\begin{equation}\label{eq:zr}
E_{ZR}=\frac{1}{4}t_{3}\int d^{3}r \rho_{n}(r)\rho_{p}(r)\rho(r) \,,
\end{equation}
where $t_{3}$ is the interaction strength for three-body force, $\rho_{n}$
($\rho_{p}$) is neutron(proton) density, and $\rho$ is total density.\\ 
The energy functional for the Coulomb interaction has an exchange term which is absent
 in classical physics,
\begin{equation}\label{eq:cou}
\begin{aligned}
E_{C} & = E_{C}^{pp} + E_{C}^{ex} \\
      & = \frac{e^{2}}{2}\int\int d^{3}r_{1} d^{3}r_{2} \frac{\rho_{p}(r_{1})
 \rho_{p}(r_{2})}{r_{12}}  -\frac{3}{4\pi}(3\pi^2)^{1/3}e^{2}\int d^{3}r 
\rho_{p}^{4/3}(r)\,.
\end{aligned}
\end{equation}
In bulk nuclear matter, the spin-orbit and Coulomb interactions constitute a
 small portion of the total energy, so we neglect these two terms in bulk 
nuclear matter.
 Then the bulk density functional would be
\begin{equation}\label{eq:bulk}
\begin{aligned}
E_{B} & = T_{kin} + E_{FR} + E_{ZR} \\
      & = \int d^{3}r  \mathcal{E}_{B}(r) \,,
\end{aligned}
\end{equation}
where $\mathcal{E}_{B}$ is the energy density for bulk matter. 
Using eq. (\ref{eq:kin}), (\ref{eq:fr}), and (\ref{eq:zr}), we find
\begin{equation}
\begin{aligned}
\mathcal{E}_{B} & = \frac{\hbar^{2} \tau_{n}}{2m}
 + \frac{\hbar^{2} \tau_{p}}{2m} \\
                & + V_{1L}(\rho_{n}\langle\rho_{n}\rangle + \rho_{p}
\langle\rho_{p}\rangle) + V_{1U}(\rho_{n}\langle\rho_{p}\rangle 
+ \rho_{p}\langle\rho_{n}\rangle) \\
                & + V_{2L}(\rho_{n}^{1+\epsilon}\langle\rho^{1+\epsilon}_{n}
\rangle + \rho_{p}^{1+\epsilon}\langle\rho^{1+\epsilon}_{p}\rangle)  + V_{2U}
(\rho_{n}^{1+\epsilon}\langle\rho^{1+\epsilon}_{p}\rangle + \rho_{p}^{1+\epsilon}
\langle\rho^{1+\epsilon}_{n}\rangle ) \\
                & + V_{3L}(\rho_{n}\langle\tau_{n}\rangle + \tau_{n}
\langle\rho_{n}\rangle + \rho_{p}\langle\tau_{p}\rangle
 + \tau_{p}\langle\rho_{p}\rangle ) + 
V_{3U}(\rho_{n}\langle\tau_{p}\rangle 
+ \tau_{n}\langle\rho_{p}\rangle + \rho_{p}
\langle\tau_{n}\rangle + \tau_{p}\langle\rho_{n}\rangle ) \\
               & + \frac{1}{4}t_{3}\rho\rho_{n}\rho_{p}
\end{aligned}
\end{equation}
where we defined the Gaussian-type integral using `$\langle ...\rangle$':
\begin{equation}
\langle u(r_{1})\rangle=\frac{1}{\pi^{3/2}r_{0}^{3}}
\int d^{3}r_{2} e^{-r^{2}_{12}/r_{0}^{2}} u(r_{2})\,.
\end{equation}

\subsection{Effective mass, potential, and thermodynamic properties}
The effective mass of nucleons at the nuclear saturation density is about
 $0.7m_{B} (m_B=938)$MeV. Some nuclear density functionals use the effective mass
 $m^{*}=m_{B}$; we, however, introduced the momentum-dependent interaction, which
 describes the effective mass of nucleons. The functional derivative $\delta E_{B}
$ gives us the effective masses and potentials in the nuclear density functional,
\begin{equation}\label{eq:gaussed}
\delta E_{B} = \int d^{3}r  \biggl( V_{n}\delta \rho_{n}
 + \frac{\hbar^{2}}{2m_{n}^{*}}\delta \tau_{n} + V_{p}\delta \rho_{p} 
+ \frac{\hbar^{2}}{2m_{p}^{*}}\delta \tau_{p} \biggr)\,.
\end{equation}
Now we get the effective mass for neutrons and protons,
\begin{equation}\label{eq:effectivemass}
m_{t}^{*} = \frac{m}{1+ 4m(V_{3L}\langle\rho_{t}\rangle
 + V_{3U}\langle\rho_{t^{\prime}}\rangle)/\hbar^{2}} \,
\end{equation}
and the potentials,
\begin{equation}
\begin{aligned}
V_{t} = & 2\biggl[V_{1L}\langle\rho_{t}\rangle +V_{1U}\langle\rho_{t^{\prime}}
\rangle+ (1+\epsilon)\rho_{t}^{\epsilon}(V_{2L}\langle\rho_{t}^{1+\epsilon}
\rangle + V_{2U}\langle\rho_{t^{\prime}}^{1+\epsilon}\rangle  )+ V_{3L}\langle
\tau_{t}\rangle + V_{3U}\langle\tau_{t^{\prime}}\rangle\biggr] \\
 & +\frac{1}{4}t_{3}(2\rho_{t}+\rho_{t^{\prime}})\rho_{t^{\prime}} \,.
\end{aligned}
\end{equation}
where $t^{\prime}$ is a different type of nucleon from $t$ nucleon.\\
The thermodynamic properties are extremely important for describing the properties of hot,
 dense matter. The degeneracy parameter in the Fermi-Dirac
 distribution function is the key to the thermodynamic properties. Using
 Fermi-integrals, we get the baryon number density and kinetic density in terms of the
 degeneracy parameter $\phi_{t}=(\mu_{t}-V_{t})/T$,
\begin{equation}\label{eq:density}
\rho_{t}=\frac{1}{2\pi^{2}}\biggl(\frac{2m_{t}^{*}T}{\hbar^{2}}\biggr)^{3/2}
F_{1/2}(\phi_{t}), \quad \tau_{t}=\frac{1}{2\pi^{2}}
\biggl(\frac{2m_{t}^{*}T}{\hbar^{2}}\biggr)^{5/2}F_{3/2}(\phi_{t})\,.
\end{equation}
Landau's quasi-particle formula gives us the entropy density $S_{t}$, 
which tells us how to find the pressure in this density functional,
\begin{equation}
S_{t}=-\frac{2}{\hbar^{3}}\int d^{3}p \big[f_{t}\ln f_{t} + (1-f_{t})
\ln (1-f_{t}) \big] =\frac{5\hbar^{2}}{6m_{t}^{*}T}\tau_{t}
 - \frac{\mu_{t}-V_{t}}{T}\rho_{t}\,.
\end{equation}
From the thermodynamic identity and entropy density given above, we can get
the pressure:
\begin{equation}
\begin{aligned}
p & = \mu_{n}\rho_{n} + \mu_{p}\rho_{p} + T S_{n} + T S_{p} -\mathcal{E} \\
  & = \sum_{t} \biggl(\frac{5\hbar^{2}}{6m_{t}^{*}}\tau_{t}
 + V_{t}\rho_{t}\biggr) -\mathcal{E} \\
  & = \frac{\hbar^{2} \tau_{n}}{3m} + \frac{\hbar^{2} \tau_{p}}{3m}
 + V_{1L}(\rho_{n}\langle\rho_{n}\rangle + \rho_{p}\langle\rho_{p}\rangle)
 + V_{1U}(\rho_{n}\langle\rho_{p}\rangle + \rho_{p}\langle\rho_{n}\rangle) \\
                &\quad + V_{2L}(1+2\epsilon)(\rho_{n}^{1+\epsilon}
\langle\rho^{1+\epsilon}_{n}\rangle + \rho_{p}^{1+\epsilon}
\langle\rho^{1+\epsilon}_{p}\rangle)  + V_{2U}(1+2\epsilon)
(\rho_{n}^{1+\epsilon}\langle\rho^{1+\epsilon}_{p}\rangle
 + \rho_{p}^{1+\epsilon}\langle\rho^{1+\epsilon}_{n}\rangle ) \\
                &\quad + V_{3L}(\rho_{n}\langle\tau_{n}\rangle 
+ \frac{7}{3}\tau_{n}\langle\rho_{n}\rangle + \rho_{p}\langle\tau_{p}
\rangle + \frac{7}{3}\tau_{p}\langle\rho_{p}\rangle) + V_{3U}(\rho_{n}
\langle\tau_{p}\rangle + \frac{7}{3}\tau_{n}\langle\rho_{p}\rangle 
+ \rho_{p}\langle\tau_{n}\rangle + \frac{7}{3}\tau_{p}\langle\rho_{n}\rangle ) \\
                &\quad + \frac{1}{2}t_{3}\rho\rho_{n}\rho_{p} \,,
\end{aligned}
\end{equation}
where in zero-temperature, non-uniform matter, the chemical potential of the protons
 and neutrons are given by,
\begin{align}\label{eq:chem}
\mu_{t} & = \frac{\hbar^{2}}{2m_{t}^{*}}(3\pi^{2}\rho_{t})^{2/3}
 + V_{t} \notag \\
        & = \frac{\hbar^{2}}{2m}(3\pi^{2}\rho_{t})^{2/3} 
 + 2(V_{3L}\langle\rho_{t}\rangle+V_{3U}\langle\rho_{t^{\prime}}\rangle)
(3\pi^{2}\rho_{t})^{2/3} \\
        &\quad + 2\biggl[V_{1L}\langle\rho_{t}\rangle +V_{1U}
\langle\rho_{t^{\prime}}\rangle+ (1+\epsilon)\rho_{t}^{\epsilon}
(V_{2L}\langle\rho_{t}^{1+\epsilon}\rangle
 + V_{2U}\langle\rho_{t^{\prime}}^{1+\epsilon}\rangle  )
+ V_{3L}\langle\tau_{t}\rangle 
+ V_{3U}\langle\tau_{t^{\prime}}\rangle\biggr] \notag\\
        &\quad+\frac{1}{4}t_{3}(2\rho_{t}+\rho_{t^{\prime}})
\rho_{t^{\prime}}\notag \,.
\end{align}

\section{Parameters for the Gaussian nuclear density functional}
Every nuclear model should reproduce five nuclear matter properties:
 binging energy, pressure, nuclear incompressibility, symmetry energy and
 effective mass, $m^{*}$. 
We use the saturation properties of nuclear matter to determine the 
parameters of the density functional. For zero-temperature, uniform nuclear 
matter, we have the energy density as a function of $u=\rho/\rho_{0}$ 
and $x=\rho_{p}/\rho$,
\begin{align}
\frac{\mathcal{E}_{B}}{T_{0}\rho_{0}}& =\frac{3}{5}2^{2/3}u^{5/3}
\big[(1-x)^{5/3}+x^{5/3}\big] + u^{2}\big[v_{1L}(x^{2}+(1-x)^{2})
 + 2v_{1U}x(1-x)\big]\notag \\
 & + 2^{1+2\epsilon}u^{2+2\epsilon}\big[v_{2L}(x^{2+2\epsilon}
 +(1-x)^{2+2\epsilon}) 
 + 2v_{2U}x^{1+\epsilon}(1-x)^{1+\epsilon} \big] \label{eq:bulkenergy}\\
 &+ 2^{2/3}u^{8/3}\big[v_{3L}(x^{8/3} +(1-x)^{8/3}) + v_{3U}(x(1-x)^{5/3}
  + x^{5/3}(1-x)) \big]+ \frac{1}{4}t_{3}^{\prime}u^{3}x(1-x) \,,\notag
\end{align}
where we define the parameters
\begin{equation}\label{eq:newdef}
\begin{aligned}
&T_{0} =\frac{\hbar^2}{2m}(3\pi^2\rho_{0}/2)^{3/2}\,, \quad v_{1L,U} 
= \frac{\rho_{0}}{T_{0}}V_{1L,U}\,, \quad v_{2L,U} = \frac{1}{T_{0}}
\Bigl(\frac{\rho_{0}}{2}\Bigr)^{1+2\epsilon} V_{2L,U}\,\\[8pt]
&v_{3L,U}  = \frac{4\cdot3^{5/3}\pi^{4/3}}{5T_{0}} 
\Bigl(\frac{\rho_{0}}{2}\Bigr)^{5/3} V_{3L,U}\, ,\quad t_{3}^{\prime} = 
\frac{\rho_{0}^{2}}{T_{0}}t_{3} \,.
\end{aligned}
\end{equation}
Now we assume that the momentum-dependent interaction is blind to 
the type of nucleon,
 so $V_{3L}=V_{3U} (v_{3}=v_{3L}+v_{3U})$. The binding enery of symmetric
 nuclear matter ($u=1$, $x=1/2$) is then given by
\begin{equation}\label{eq:unidensity}
\frac{\mathcal{E}_{0}}{\rho_{0}}=-B_{0}=T_{0}\biggl[\frac{3}{5}
 + \frac{v_{1L} + v_{1U}}{2} + v_{2L} + v_{2U} + \frac{v_{3}}{2}
 + \frac{t_{3}^{\prime}}{16}\biggr]\,,
\end{equation}
where $B_{0}=16$MeV is the binding energy per baryon at the nuclear saturation
 density.\\
The pressure at the saturation density vanishes, which mean the enery per baryon
 has its minimum at the saturation density,
\begin{equation}\label{eq:unipress}
p_{0}=\rho_{0}T_{0}\biggl[\frac{2}{5} + \frac{v_{1L} + v_{1U}}{2} 
+ (1+2\epsilon)(v_{2L}+v_{2U}) + \frac{5}{6}v_{3} + \frac{1}{8}t_{3}^{\prime}
 \biggr]=0\,.
\end{equation}
The incompressibility parameter at the saturation density is
 given by
\begin{equation}\label{eq:incompress}
\begin{aligned}
K_{0}& =9\frac{dp}{d\rho}\bigg{\vert}_{\rho=\rho_{0}}=T_{0}\big[6 + 9(v_{1L} 
+ v_{1U}) + 9(1+2\epsilon)(2+2\epsilon)(v_{2L}+v_{2U}) + 20v_{3 }
 + \frac{27}{8}t_{3}^{\prime}\big]\\
 &=265 \text{MeV}\,.
\end{aligned}
\end{equation}
The symmetry energy in nuclear matter is defined as
\begin{equation}\label{eq:symmetry}
\begin{aligned}
S_{v} & =\frac{1}{8}\frac{d^{2}(\mathcal{E}/\rho)}{d x^{2}}
\bigg{\vert}_{\rho=\rho_{0}, x=1/2} \\[8pt]
      & = T_{0}\biggl[\frac{1}{3} + \frac{v_{1L}-v_{1U}}{2} 
 + (1+\epsilon)((1+2\epsilon)v_{2L}-v_{2U}) + \frac{5}{18}v_{3}
 -\frac{1}{16}t_{3}^{\prime}\biggr] \\
 & = 28 \text{MeV}\,.
 \end{aligned}
\end{equation}
Another parameter, which is related to symmetry energy, is given by
\begin{equation}
\begin{aligned}\label{eq:l}
L & =\frac{3\rho_{0}}{8}\frac{d^{3}(\mathcal{E}/\rho)}{d\rho dx^{2}}
\bigg{\vert}_{\rho=\rho_{0}, x=1/2} \\[8pt]
  & = T_{0}\biggl[\frac{2}{3} + \frac{3}{2}(v_{1L}-v_{1U})
 + 3(1+2\epsilon)(1+\epsilon)((1+2\epsilon)v_{2L}-v_{2U}) + \frac{25}{18}v_{3}
 -\frac{3}{8}t_{3}^{\prime}\biggr]\\
 &=54\text{MeV}\,.
\end{aligned}
\end{equation}
We choose the effective mass at the saturation density as 0.78$m_{b}$ and use
 this number in the eq. (\ref{eq:effectivemass}) :
\begin{equation}
m^{*}=\frac{m}{1+ 2m\rho_{0}V_{3}/\hbar^{2}}=0.78m_{b}\,.
\end{equation}
Thus we can easily recover $v_{3}$ from eq. (\ref{eq:newdef}). 
From eq. (\ref{eq:unidensity}), (\ref{eq:unipress}), and (\ref{eq:incompress}), 
we can have $v_{1}=v_{1L}+v_{2L}$, $v_{2}=v_{2L}+v_{2U}$ and $t_{3}^{\prime}$. 
\begin{equation}
\begin{aligned}
v_{1} & = \frac{5K_{0}/T_{0} + 5v_{3}(1-3\epsilon)
 -72\epsilon - 90B_{0}/T_{0}(1+2\epsilon) -12 }{45\epsilon}\\[8pt]
v_{2} & = \frac{12 + 90B_{0}/T_{0} -5K_{0}/T_{0} -5v_{3}}
{90\epsilon(1-2\epsilon)} \\[8pt]
t_{3}^{\prime} & = -8v_{1} -16v_{2}-8v_{3} -\frac{16B_{0}}{T_{0}} -\frac{48}{5}\,.\\
\end{aligned}
\end{equation}
Then we can manipulate eq. (\ref{eq:symmetry}), and (\ref{eq:l}) to get $v_{1L}$
 and $v_{2L}$,
\begin{equation}
\begin{aligned}
v_{1L} & = \frac{1}{2}v_{1} +\frac{1}{2\epsilon}
\biggl[\frac{5(1-3\epsilon)}{27}v_{3} + \frac{2\epsilon -1}{16}t_{3}^{\prime}
 + \frac{(1+2\epsilon)S_{v}}{T_{0}} -\frac{L}{3T_{0}}
 -\frac{1+6\epsilon}{9}\biggr]\\[8pt]
v_{2L} & = \frac{1}{2(1+\epsilon)}v_{2}
 -\frac{1}{4\epsilon(1+\epsilon)^{2}}\biggl[\frac{5}{27}v_{3}
 -\frac{t_{3}^{\prime}}{16} + \frac{S_{v}}{T_{0}} -\frac{L}{3T_{0}}
 -\frac{1}{9} \biggr] \,,\\
\end{aligned}
\end{equation}
and we can have $v_{1U}=v_{1}-v_{1L}$ and $v_{2U}=v_{2}-v_{2L}$. 

\subsection{Determination of $1+\epsilon$ power}
We added in eq (\ref{eq:fr}) the auxiliary density interaction 
with the $1+\epsilon$ power. We might
 regard $1+\epsilon$ as the many-body effect---for example, a three-body force if
 $\epsilon > \frac{1}{2}$. It is known, however, that interactions among  
 more than three-bodies are unimportant in dense matter. Thus we might restrict
  the $\epsilon$ to be less
 than $\frac{1}{2}$. As the $\epsilon$ changes, the $t_{3}$ parameter changes
 sign which means the three-body force can be attractive or repulsive. In the general
 Skyrme model with the three-body force, the $t_{3}$ parameter is positive. We
 choose $\epsilon=1/6$ so that the interaction has the form of
 $\rho_{t_{1}}^{7/6}\rho_{t_{2}}^{7/6}$. In zero-temperature, uniform matter,
 we have $u^{7/3}$ terms in the energy density. 
From eq. (\ref{eq:bulkenergy}), the energy density has
 $u^{5/3}$, $u^{2}$, $u^{7/3}$, $u^{8/3}$, and $u^{3}$ terms if we have
 $\epsilon = 1/6$ so we can use a statistical approach in uniform matter.
\begin{table}
\begin{center}
\caption{ Interaction parameters when $\epsilon=1/6$, $K=265$ MeV, $S_{v}=28$MeV, 
$L=54$MeV.}
\begin{tabular}{c c c c c c c c c c c c c}\hline\hline
$v_{1L}$ && $v_{1U}$ && $v_{2L}$ && $v_{2U}$ && $v_{3L,U}$ && $t_{3}^{\prime}$ && $r_{0}$ (fm)\\ \hline
-1.766 && -3.472 &&  0.410 && 0.931 && 0.169 && 1.177 && 1.205 \\ \hline
\end{tabular}
\end{center}
\end{table}
Fig 1. shows the energy per baryon from GNDF and APR\cite
{apr1998} EOS. We can see that as the density increases, the pressure from the
 two models agrees very well.
\begin{center}
  \begin{figure}[h]
    \scalebox{0.6}{\includegraphics{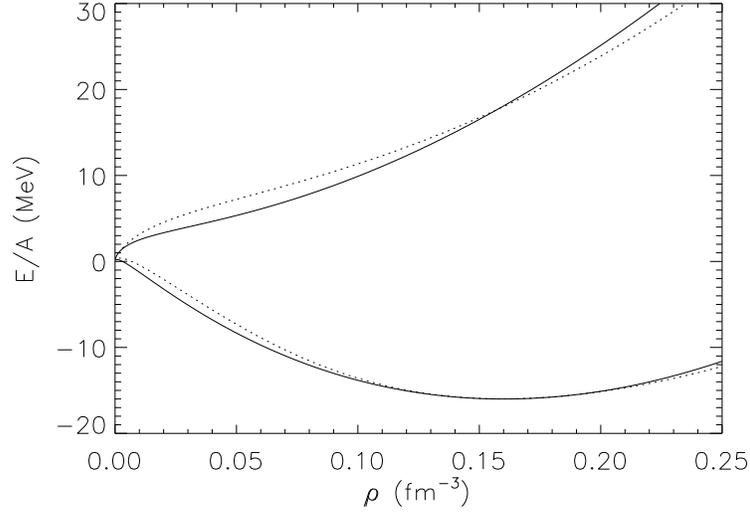}}
  \caption{The solid line represents the enery per baryon (uniform matter) using
 GNDF. The upper (lower) curve represents the energy per baryon of
 pure neutron matter(symmetric nuclear matter).  The energy per baryon (dotted line) 
 from the APR\cite{apr1998} EOS was added for comparison. }
  \end{figure}
\end{center}

\subsection{The effective range of the nuclear force : $r_{0}$}
In the Gaussian-interaction model, we can see the effective range of the force
 is given by $r_{0}$, which is approximately $\sim $ 1 to 2 fm. We don't have an
 analytic form of $r_{0}$, so we need to rely on the numerical solution of the
 surface tension of semi-infinite nuclear matter:
\begin{equation}
\omega =\int_{-\infty}^{\infty} \big[\mathcal{E} -TS_{n} -TS_{p}
 -\mu_{n}\rho_{n} -\mu_{p}\rho_{p} + p_{0} \big]dz =
 -\int_{-\infty}^{\infty}\big[p(z) - p_{0}\big] dz\,,
\end{equation}
where $p_{0}$ is the pressure at $z= -\infty$ or $z=+\infty$.
In one-dimensional, semi-infinite nuclear matter, we assume that the nuclear
 density depends only on the z-axis; the Gaussian integral then becomes
\begin{equation}
\frac{1}{\pi^{3/2}r_{0}^{3}}\int d^{3}r \, u(r)
 = \frac{1}{\pi^{1/2}r_{0}} \int_{-\infty}^{+\infty} dz \,u(z)\,.
\end{equation}
Experimental values of surface tension and surface thickness are 
$\omega=1.250$MeV fm$^{-2}$ and $t_{90-10}=2.3$fm. Fig.2 shows the numerical 
calculation, which says that $r_{0}=1.205$fm from the surface tension 
and $r_{0}=1.149$
 from $t_{90-10}$ thickness. There is a 5$\%$ discrepancy between the two results.
 Table 1 shows the interaction parameters which we use in this paper 
 when $K=265$MeV, $S_{v}=28$MeV, $L=54$MeV, and $\epsilon=1/6$. The simple 
 density dependent interactions ($v_{1L,U}$) are attractive, on the other hand,
 the auxiliary density dependent interactions ($v_{2L,U}$),
 momentum dependent ineractions ($v_{3L,U}$)
 three-body force($t_{3}$) are repulsive
 in our model.
 
\begin{figure}\label{findingr}
\begin{center}
\begin{tabular}{cc}
\scalebox{0.45}{\includegraphics{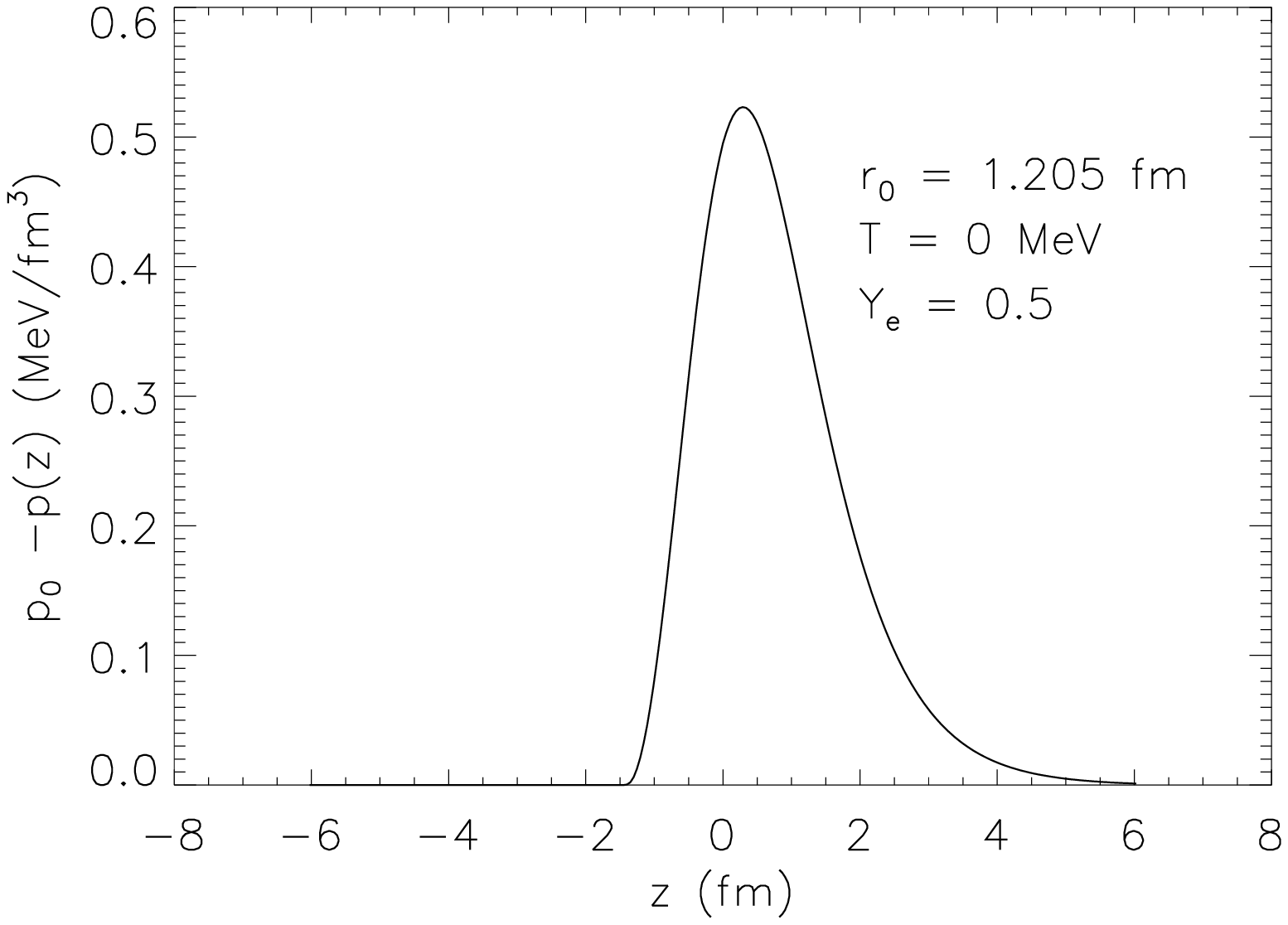}}
&\scalebox{0.45}{\includegraphics{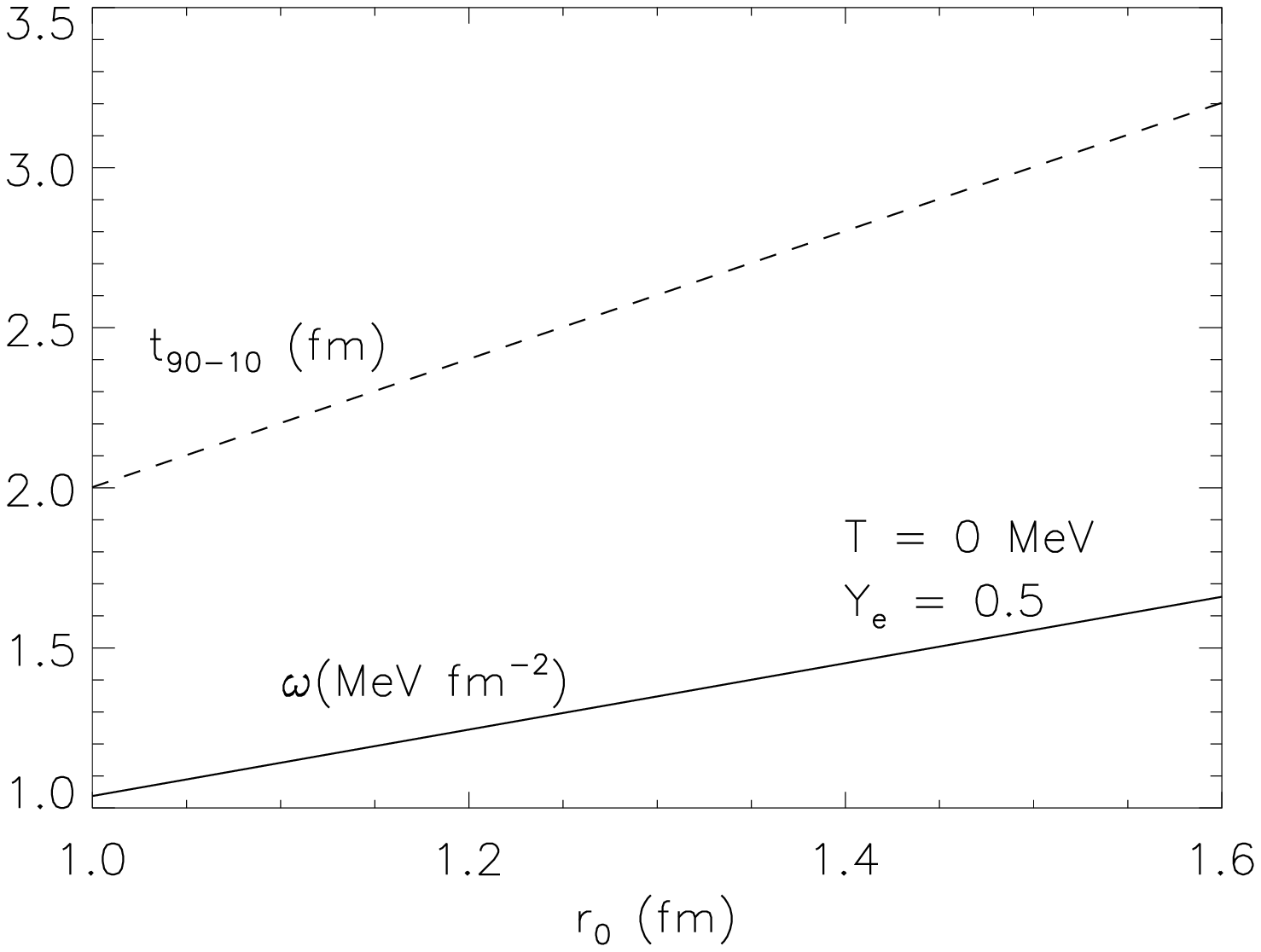}}
\end{tabular}
\caption{The left figure shows the quantity $p_{0}-p(z)$ at the semi-infinite
 nuclear surface when $r_{0}=1.205$fm and Y$_{e}=0.5$. The surface tension from
 this configuration is $\omega=1.250$MeV fm$^{-2}$. The right figure shows 
 the surface tension (solid line) and $t_{90-10}$ thickness (dashed line) as a
 function of $r_{0}$. When $r_{0}=1.205$, $t_{90-10}=2.412$fm.
 The surface tension and $t_{90-10}$ thickness are both linear functions
  of $r_{0}$.}
\end{center}
\end{figure}

\section{Nuclear matter and nuclei}
\subsection{Specific heat}
The specific heat of uniform nuclear matter can be obtained by
\begin{equation}
C_{V}=T\frac{\partial S}{\partial T}\bigg{\vert}_{\rho}
=\frac{\partial E}{\partial T}\bigg{\vert}_{\rho}\,.
\end{equation}
For a non-interacting Fermion gas, the specific heat increases linearly with 
temperature. When the temperature is low enough, we expect 
that the specific heat of the nuclear matter tends to behave like a free Fermion gas.
 The specific heat formula for degenerate gas is given by\cite{jhbook}
\begin{equation}\label{eq:specheat}
C_{V}=\frac{1}{3}m^{*}k_{F}k_{B}^{2}T\,,
\end{equation}
where $m^{*}$ is effective mass of a nucleon, $k_{F}$ is the Fermi momentum, and 
$k_{B}$ is Boltzman constant.\\
However, as the temperature increases, the non-linear behavior of the specific
 heat comes out so the degenerate gas formula is no longer valid, since the
 nucleons deep inside the Fermi surface are excited.
\begin{center}
  \begin{figure}[h]
    \scalebox{0.6}{\includegraphics{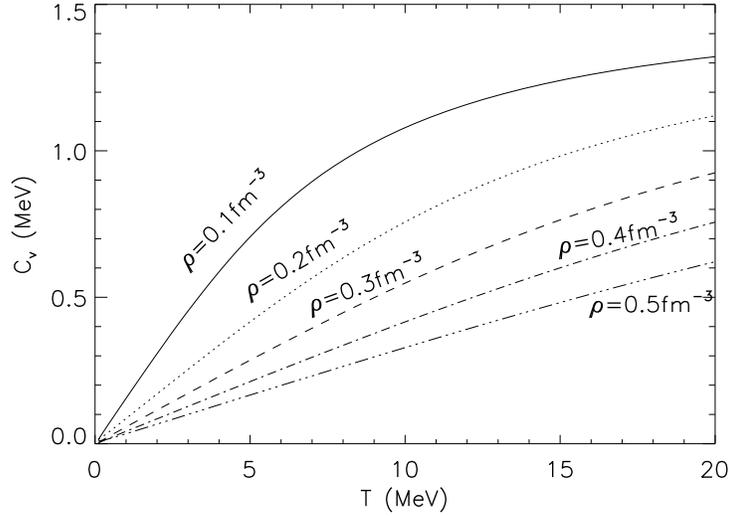}}
  \caption{This figure shows the specific heat per nucleon of uniform matter
 for different densities. If the temperature is low enough, the specific heat
 behaves linearly with temperature.}
  \end{figure}
\end{center}
To calculate the specific heat of uniform nuclear matter, we use the Johns, 
Ellis, and Lattimer (JEL) method \cite{jel1996}, 
which enables us to get the pressure, energy density and entropy density 
for a given degeneracy parameter. Fig.3 shows the specific heat per nucleon 
of uniform nuclear matter. It shows the linear relation between the specfic heat and
temperature at low temperation as in eq. (\ref{eq:specheat}).\\
A detailed calculation of the specific heat at
 sub nuclear density in the neutron star needs to take into account the beta-equilibrium
 condition and heavy nuclei with a neutron gas. The specific heat plays an
 important role in the cooling process of neutron stars. In the neutron star 
 crust, there are heavy nuclei and a free-neutron gas. The effective masses of protons
 and neutrons are different from the center of the heavy nuclei and dilute neutron
gas, so we can't use eq. (\ref{eq:specheat}).  In this case the specific heat at the
 neutron star crust can be calculated numerically by changing temperature and
 comparing the total energy change.

\subsection{Nuclei at $T=0$ MeV}
We can use the GNDF theory and the Lagrange multiplier method to find the radius and binding 
energy per nucleon for a single nucleus using the Winger-Seitz cell method. In 
the Lagrange
multiplier method, the chemical potentials of protons and neutrons are constant 
in the Wigner-Seitz cell to minimize the total free energy. Fig. 4
 shows the radius and binding energy of the closed-shell nuclei using
this method. These results
 agree well with experiment\cite{st2005}. $^{40}$Ca has a larger charge 
 radius than
 neutron radius because of the Coulomb repulsion between protons. 
 The solid (dotted) line denotes neutron (proton) density. As the atomic number 
 increases, the central density of neutrons increases; on the other hand the 
 central density of
protons decreases. The difference between charge and neutron radii increases and the 
neutron skin becomes thicker as the atomic number increases. In $^{208}$Pb nuclei, 
the central density of protons is lower than the proton density of the outer
 part of nuclei  ($r= 4 - 5$ fm) because of proton Coulomb repulsion.
\begin{figure}\label{closenuclei}
\begin{center}
\begin{tabular}{ccc}
\scalebox{0.35}{\includegraphics{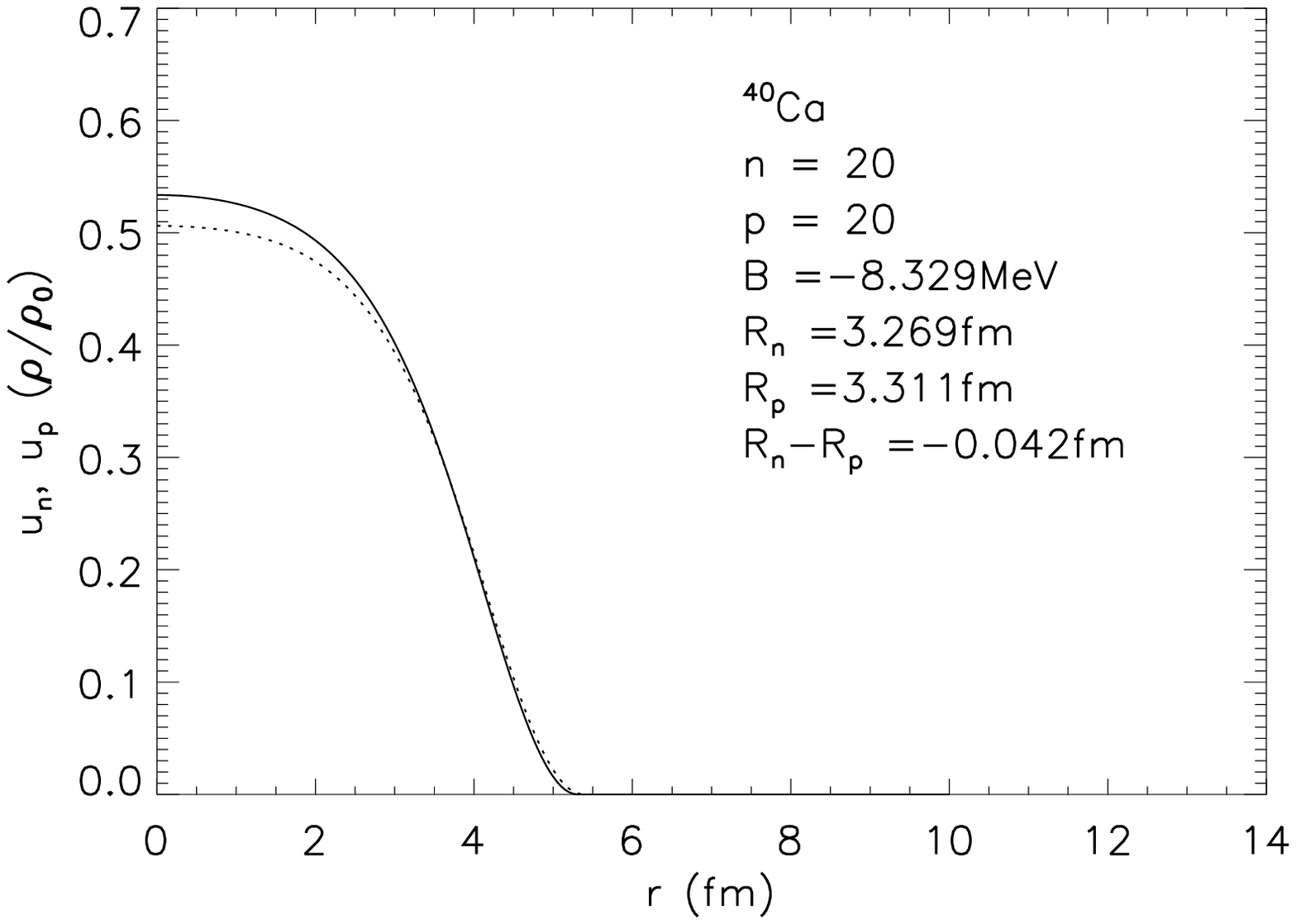}}&
\scalebox{0.35}{\includegraphics{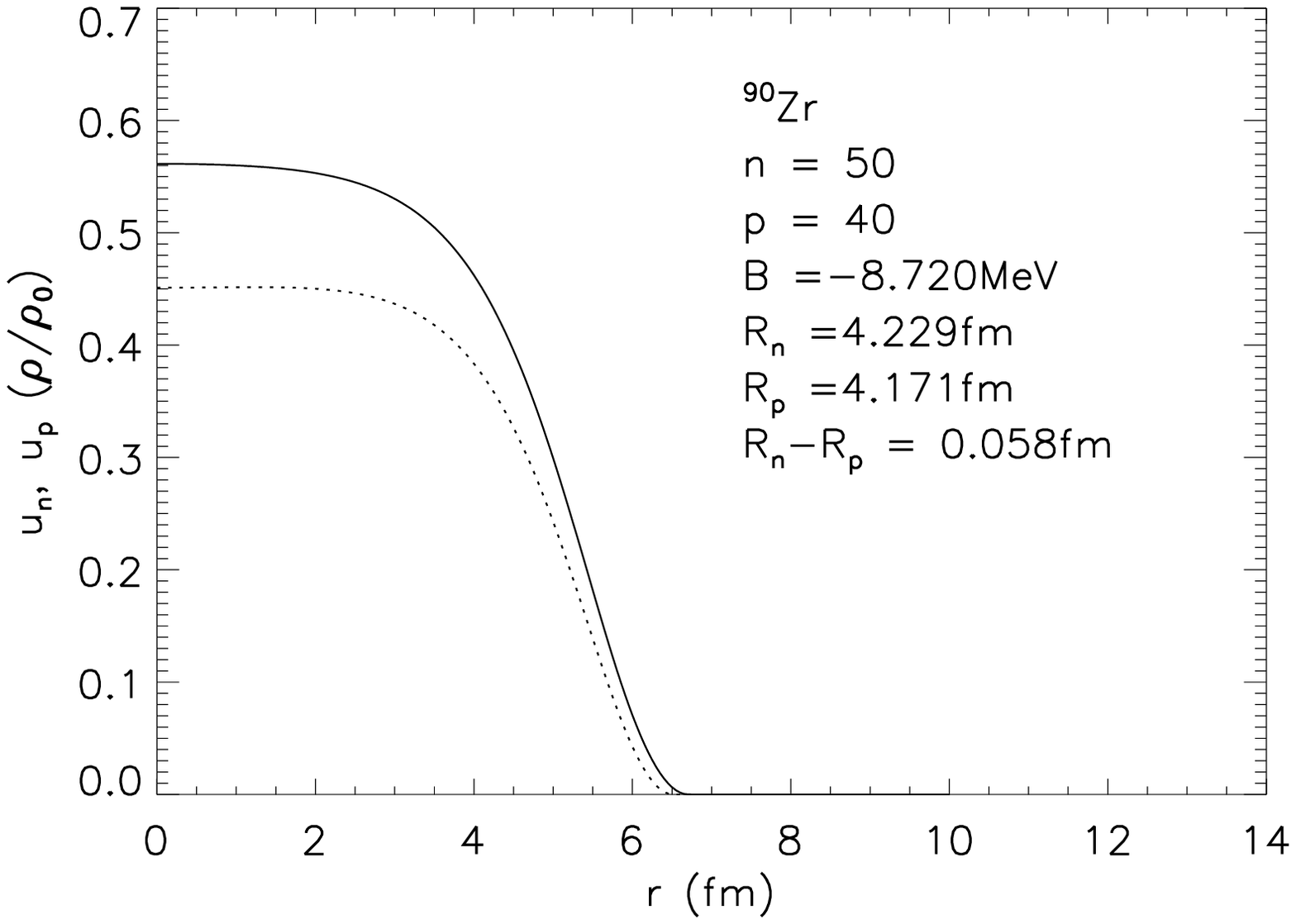}}&
\scalebox{0.35}{\includegraphics{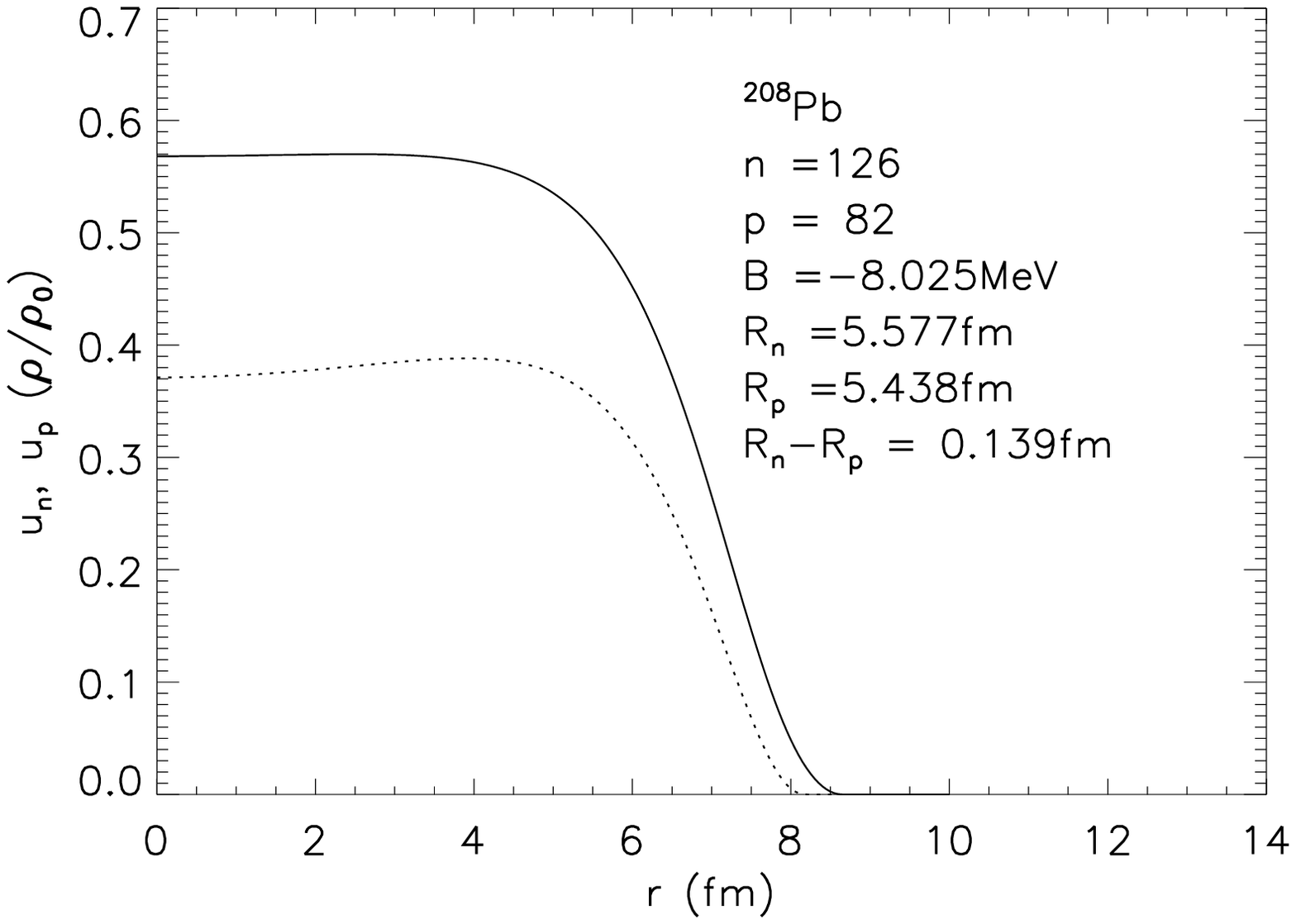}}
\end{tabular}
\caption{This figure shows the basic properties of the closed-shell nuclei which 
can be obtained from the GNDF. The solid (dotted) line indicates the neutron (proton)
 density as a function of radius. As the number of nucelons
 in the nuclei increase, the neutron skin thickness($R_{n}-R_{p}$) increases. }
\end{center}
\end{figure}

\begin{table}[h]%[h!b!p!]
\caption{Comparsion of the results from Steiner (Potential $\&$ Field Theoretical)
 et al.\cite{st2005} , FRTF I\cite{frtf1}, FRTF II\cite{frtf2} and the GNDF}
\begin{tabular}{cccccccc}%{lllllll}
\hline
% \T and \B would not work if it is placed here (needs to go inside cell)
Nucleus &  Property & Experiment & Potential & FT & FRTF I &FRTF II& GNDF \\
\hline
$^{208}$Pb  & $ r_{ch}$ (fm)  & 5.50  & 5.41 & 5.41 & 5.38 &5.45& 5.44  \\
                        & BE/A(MeV)    & 7.87  & 7.87 & 7.77 & 8.01 &8.17& 8.02 \\
                        & $\delta$R(fm)& 0.12 $\pm$ 0.05 & 0.19 & 0.20 & 0.15 &0.13& 0.14 \\
                        &              & 0.20 $\pm$ 0.04  &      &      &      &&   \\[7pt]
$^{90}$Zr  & $r_{ch}$ (fm)  & 4.27  & 4.18 & 4.17 & 4.10  & 4.15&4.17 \\
                        & BE/A(MeV)    & 8.71   & 8.88 & 8.65 & 8.77 & 9.00&8.72 \\
                        & $\delta$R(fm)& 0.09 $\pm$ 0.07   & 0.075 & 0.093 & 0.064 & 0.054&0.057 \\[7pt]
$^{40}$Ca  & $r_{ch}$ (fm)  & 3.48  & 3.40 & 3.34 & 3.22 & 3.26&3.31 \\
                        & BE/A(MeV)    & 8.45  & 8.89 & 8.61 & 8.47 &8.77 &8.33 \\
                        & $\delta$R(fm)& -0.06 $\pm$ 0.05 & -0.044 & -0.046 & -0.036 & -0.039& -0.042 \\
                        &              & -0.05 $\pm$ 0.04  &      &      &      & &  \\
\hline
\end{tabular}
\label{table1}
\end{table}
Table II shows the proton and neutron radii and binding energy per baryon of 
closed shell nuclei from various nuclear models. The calculation from GNDF
theory agrees well with experimental results.

\subsection{Heavy nuclei in the neutron star crust}
In the neutron star crust, heavy nuclei are formed with a free-neutron gas.
These heavy nuclei are suspected to form a BCC (body centered cubic) structure.
In the static equilibrium state, we calculate the density profile of heavy 
nuclei with a neutron gas using the Wigner-Seitz Cell method. 
The plot on the left side of Fig. 5 shows
 the proton (dotted line) and neutron density (solid line) 
profiles from the center (r=0 fm) of heavy nuclei when $\rho=0.01$fm$^{-3}$. 
There are dripped neutrons outside of the heavy nuclei. 
The cell size (Rc), which is a rough estimate of the distance between
neighboring heavy nuclei, is determined by nuclear density and beta 
equilibrium conditions ($\mu_{n}=\mu_{p} + \mu_{e}$).
There is a wave function overlap at the boundary of Wigner-Setiz cell. The
actual distance between heavy nuclei is $(8\pi/3)^{1/3}$Rc.
The right side of Fig. 5 shows the binding energy per baryon 
as a function of  Wigner-Seitz cell size. 
As the density decreases, the cell size increases and 
the energy per baryon converges to -8.0 MeV. 
\begin{figure}\label{heavynuclei}
\begin{center}
\begin{tabular}{cc}
\scalebox{0.4}{\includegraphics{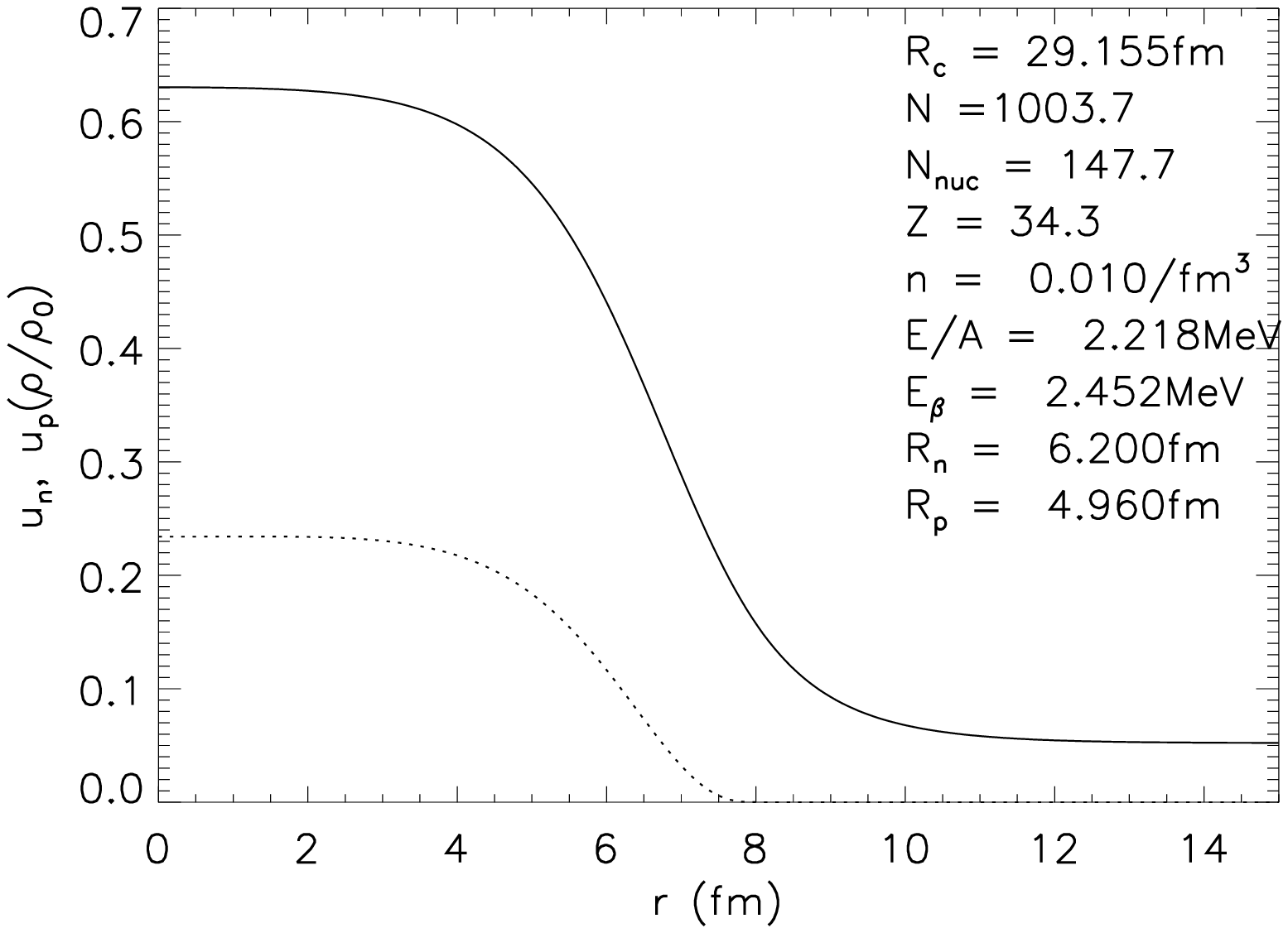}}&
  \scalebox{0.4}{\includegraphics{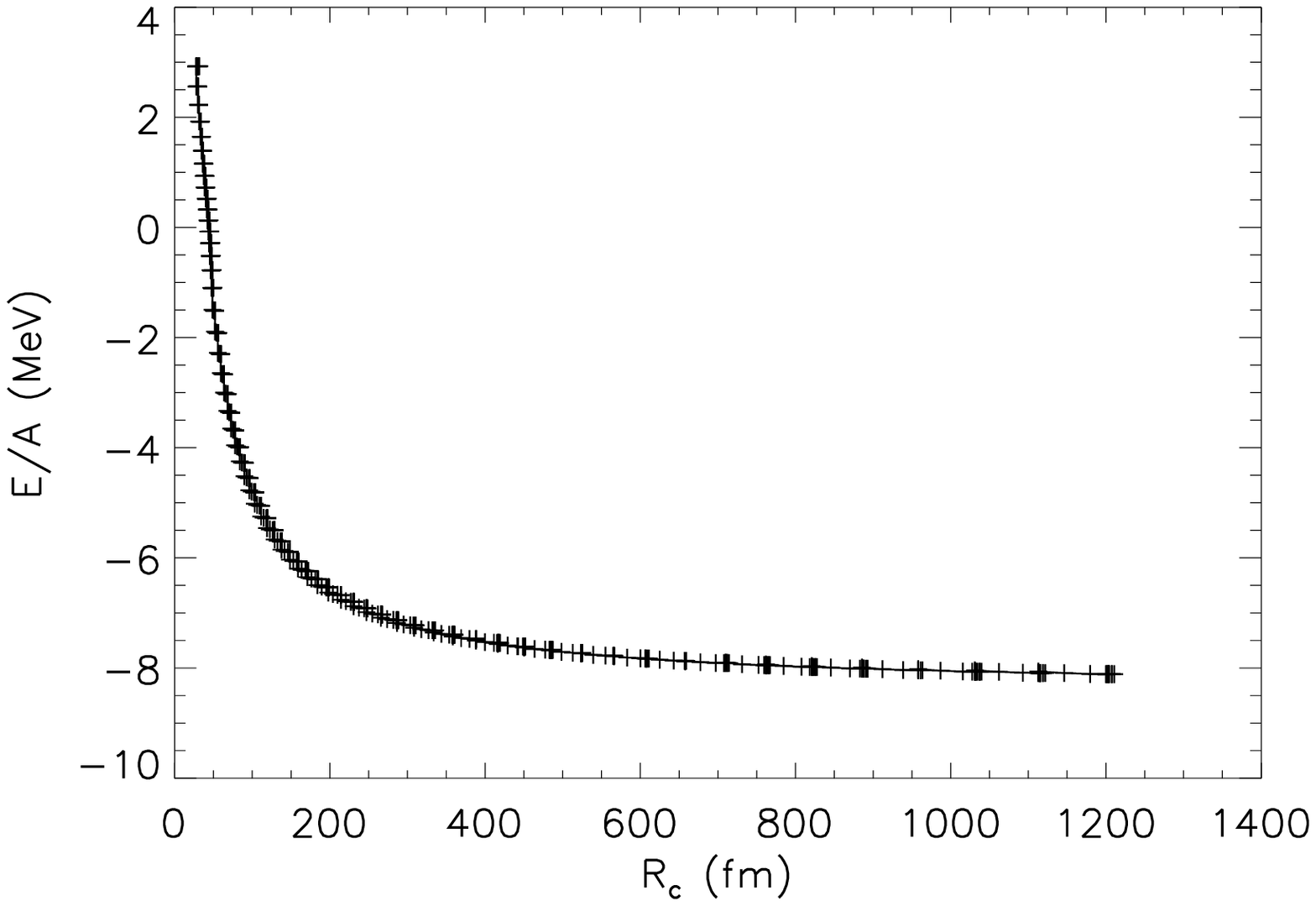}}
\end{tabular}
\caption{In a neutron star, heavy nuclei exist. The left figure shows the
density profile of proton (dotted line) and neutron (solid line). $r=0$ fm means
the center of heavy nuclei. Outside the heavy nuclei, there are dripped neutrons. 
 As density decreases, the cell size increases and the
 energy per baryon converges to -8.0 MeV.}
\end{center}
\end{figure}

\begin{table}[h]%[h!b!p!]
\caption{Nuclear properties in the neutron star crust}
\begin{tabular}{cccccccccccc}%{lllllll}
\hline
$\rho$ (fm$^{3}$) &&  $p$ (MeV/fm$^{3}$)        && $\epsilon$ (MeV/fm$^{3}$)& & N$_{\text{nuc}}$ & & Z  & & Rc (fm) \\
\hline
 5.623$\times10^{-2}$ & & 0.181                 && 53.06             &&  271.6 &&  89.51 &&  25.12 \\
 5.012$\times10^{-2}$ & & 0.147                 && 47.27             &&  218.0 &&  32.18 &&  18.70 \\
 3.981$\times10^{-2}$ & & 9.332$\times10^{-2}$  && 37.52             &&  137.8 &&  21.69 &&  18.06 \\
 2.512$\times10^{-2}$ & & 3.965$\times10^{-2}$  && 23.65             &&  145.3 &&  27.93 &&  22.77 \\
 1.585$\times10^{-2}$ & & 1.976$\times10^{-2}$  && 14.91             &&  150.7 &&  32.07 &&  26.36 \\
 1$\times10^{-2}$   & & 1.081$\times10^{-2}$  && 9.405             &&  147.7 &&  34.29 &&  29.15 \\
 1$\times10^{-3}$   & & 7.988$\times 10^{-4}$ && 9.383$\times10^{-1}$ && 116.3&& 38.50   && 40.71\\
 1$\times10^{-4}$   & & 6.945$\times 10^{-5}$ && 9.352$\times10^{-2}$ && 79.23  && 38.31   && 65.47\\
 1$\times10^{-5}$   & & 1.480$\times 10^{-6}$ && 9.326$\times10^{-3}$ && 58.02  && 36.98   && 131.4\\
 1$\times10^{-6}$   & & 2.362$\times 10^{-9}$ && 9.311$\times10^{-4}$ && 47.90  && 35.14   && 270.6\\
 1$\times10^{-7}$   & & 1.714$\times 10^{-9}$ && 9.303$\times10^{-5}$ && 43.37  && 34.10   && 569.7\\
\hline
\end{tabular}
\label{table3}
\end{table}
%% Figure of density profile heavy nuclei and effective mass of nuclei at position
%
Table III shows the thermodynamic properties and physical dimensions of nuclei 
in the neutron star crust. N$_{\text{nuc}}$ and Z are number of neutrons
 and protons of heavy nuclei in the Wigner-Seitz Cell.  
The atomic number of heavy nuclei remains Z $\sim$ 35 for a large range of 
densities before the phase transition to uniform matter. 
This means that the proton fraction decreases 
as the density increases. For a narrow range of densities, the atomic 
number suddenly increases and the heavy nuclei merge with free neutrons to make
uniform nuclear matter.
\begin{center}
  \begin{figure}[h]
    \scalebox{0.5}{\includegraphics{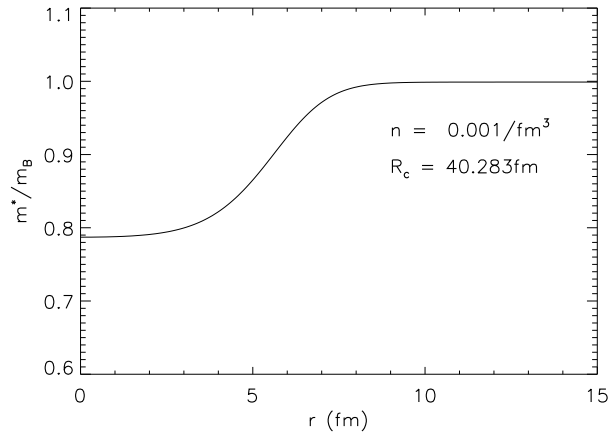}}
  \caption{Effective mass of nucleons in the Wigner-Seitz cell as a function of
  radial distance from the center of heavy nuclei. Since the nuclear 
  interaction is weak at the boundary of the cell, the effective mass of 
  nucleon and the pure mass of a nucleon become equal.}
  \end{figure}
\end{center}
Fig. 6 shows the effective mass of nucleons in the Wigner-Seitz cell. The 
effective mass of nucleons in the Wigner-Seitz cell is given by 
eq. (\ref{eq:effectivemass}) and eq. (\ref{eq:newdef}),
\begin{equation}
m^{*}_{t}=\frac{m}{1+\frac{5}{3}\left(
v_{3L}\frac{\tilde{\rho_{t}}}{\rho_{0}}
 + v_{3U}\frac{\tilde{\rho_{t^{\prime}}}}{\rho_{0}}\right)}\,.
\end{equation}
Since we assume the momentum interaction is blind with respect to isospin,
 the effective mass
is identical for different isospin nucleons. The effective mass to pure mass 
ratio of nucleons is 0.78 
at the center of the heavy nuclei and becomes 1 at the outer region of 
the Wigner-Seitz cells since the density of nuclear matter is low at
the outside of the heavy nuclei, the interaction energy of nuclear matter
is weak.

\section{Phase transition}
In the neutron star, we can see two types of phase transitions: one is the 
phase transition from nuclei with a neutron gas to uniform matter, and the other
is the phase transition from uniform nuclear matter to quark matter. During
 the first phase transition, we can see the nuclear pasta phase. That is, spherical
 nuclei become ellipsoidal, then cylindrical, and finally slab phase before nuclear
 matter becomes uniform matter. However, the energy difference is quite small,
so that the effects on the large scale physics are neglible. On the other hand, the
 second phase transition is quite dramatic. The energy and pressure change 
significantly from nuclear matter to quark matter.

\subsection{Uniform matter}
 To check the phase transition points from heavy nuclei with a neutron gas to
uniform nuclear matter, we can simply compare the energy per baryon of uniform
 nuclear matter with the energy per baryon of nuclei with a neutron gas since
the nuclear matter exists in the lowest energy states.
 The energy per baryon in uniform
 matter can be easily obtained by changing the `$\langle \dots \rangle$' integrals 
to non  integral form from eq. (\ref{eq:gaussed}) since the Gaussian 
integrations in uniform matter become unity.
Typically there is a phase transition around 0.5$\rho_{0}$.\\
We know that in the outer crust of the neutron star the nuclei has a BCC
 structure. If we assume that the pasta phase exists in the low-density region,
 we may use the density perturbation to see the phase transition from nuclei with 
a neutron gas to uniform nuclear matter. We use the wave number pertubation to 
see the energy exchange, which has contributions from the volume effects,
 gradient effects, and Coulomb energy can be approximated\cite{pr1995},
\begin{equation}
v(q) \simeq v_{0} + \beta q^{2} +\frac{4\pi e^{2}}{q^{2}+ k_{TF}^2}\,,
\end{equation}
where $q$ is the sinusoidal variation of the wave number in the spatially 
periodic density perturbation.\\
The volume term is given by
\begin{equation}\label{eq:volume}
v_{0}=\frac{\partial \mu_{p}}{\partial \rho_{p}} -\frac{(\partial \mu_{p}/
\partial \rho_{n})^{2}}{(\partial \mu_{n}/\partial \rho_{n})}\,.
\end{equation}
The energy exchange from the gradient has the form
\begin{equation}
\beta=D_{pp} + 2D_{np}\xi + D_{nn}\xi^{2}\,, \quad \xi 
=-\frac{\partial \mu_{p}/\partial \rho_{n}}{\partial \mu_{n}/\partial \rho_{n}}
\end{equation}
where the coefficients of the gradient terms are given by
$D_{pp}=D_{np}=D_{nn} =132 \text{MeV}\cdot \text{fm}^{5}$\cite{jcbh2009}.
 The $k_{TF}$ in the Coulomb interaction represents the inverse Thomas-Fermi
 screening length of the electrons. When we see the change in the sign of $v$, 
the uniform matter phase is more stable than the perodic structure of the nuclei.
 The $v$ has a minimum at
\begin{equation}
v_{min}= v_{0} + 2(4\pi e^2 \beta)^{1/2} -\beta k_{TF}^{2}\,,
\end{equation}
when $q_{min}^{2}=(4\pi e^{2}/\beta)^{1/2} -k_{TF}^{2}$.\\
Another way to see the phase transition is to use the thermodynamic instability.
 The thermodynamic stability condition can be described using the
inequalities\cite{lp2007}\cite{jcbh2009},
\begin{equation}\label{eq:stable}
\begin{aligned}
-\biggl(\frac{\partial P}{\partial v}\biggr)_{\mu} > & 0, \\[8pt]
-\biggl(\frac{\partial \mu}{\partial q_{c}}\biggr)_{v} > & 0.
\end{aligned}
\end{equation}
where $P=P_{b} + P_{e}$ is the total pressure from electrons and baryons 
and $\mu=\mu_{n}-\mu_{p}$ is the difference between the neutron and 
proton chemical potentials, which is the electron chemical potential in 
beta-stable matter. $q_{c}$ is defined as $q_{c}=x_{p}-\rho_{e}/\rho$. 
Mathematically, the inequalities in eq. (\ref{eq:stable}) show that
the energy per baryon is convex. 
Eq. (\ref{eq:stable}) can be verified to be 
\cite{lp2007}\cite{jcbh2009}
\begin{equation}\label{eq:simpl1}
\begin{aligned}
-\biggl(\frac{\partial P}{\partial v}\biggr)_{\mu} & =\rho^{2}
\biggl[2\rho \frac{\partial E(\rho, x_{p})}{\partial \rho} 
+ \rho^{2}\frac{\partial^{2} E(\rho, x_{p})}{\partial \rho^{2}}
-\biggl(\frac{\partial^{2}E(\rho, x_{p})}{\partial \rho 
\partial x_{p}}\rho\biggr)^{2}\Big{/}\frac{\partial^{2}E(\rho, x_{p})}{\partial x^{2}_{p}}
\biggr] > 0, \\[8pt]
-\biggl(\frac{\partial \mu}{\partial q_{c}}\biggr)_{v} & 
=\biggl(\frac{\partial^{2}E(\rho, x_{p})}{\partial x^{2}_{p}}\biggr)^{-1}
+ \frac{\mu_{e}^{2}}{\pi^{2}\hbar^{3}\rho} > 0\, .
\end{aligned}
\end{equation}
The second of eq. (\ref{eq:simpl1}) always holds, so the first will determine
 the phase transition in the neutron star crust. In Xu et al.\cite{jcbh2009},
 they use a simple equation to determine the instability 
using the thermodynamic relation,
\begin{equation}\label{eq:therins}
\frac{2}{\rho}\frac{\partial E}{\partial \rho}\frac{\partial^{2} E}
{\partial x_{p}^{2}} + \frac{\partial^{2}E}{\partial \rho^{2}}
\frac{\partial^{2} E}{\partial x_{p}^{2}} -\biggl(\frac{\partial^{2} E}
{\partial \rho \partial x_{p}}\biggr)^{2} =\frac{\partial \mu_{n}}
{\partial \rho_{n}}\frac{\partial \mu_{p}}{\partial \rho_{p}} 
-\biggl(\frac{\partial \mu_{n}}{\partial \rho_{p}} \biggr)^{2} \,.
\end{equation}
Eq. (\ref{eq:therins}) is equilivalent to the volume part of the 
thermodynamic perturbation eq. (\ref{eq:volume}) method except that
 there is a $\partial \mu_{n}/\partial \rho_{n}$ difference.
 Comparing the two methods (pertubation and thermodynamic instability)
 shows the effects of the gradient and Coulomb terms in the pertubation method
 on the transition densities. Fig. 7 shows transtion densities using the
perturbation method and thermodynamic instability. The perturbation method
has a lower transition density (0.355$\rho_{0}$) than thermodynamic 
instibility method (0.406$\rho_{0}$). 
\begin{center}
  \begin{figure}[h]
    \scalebox{0.6}{\includegraphics{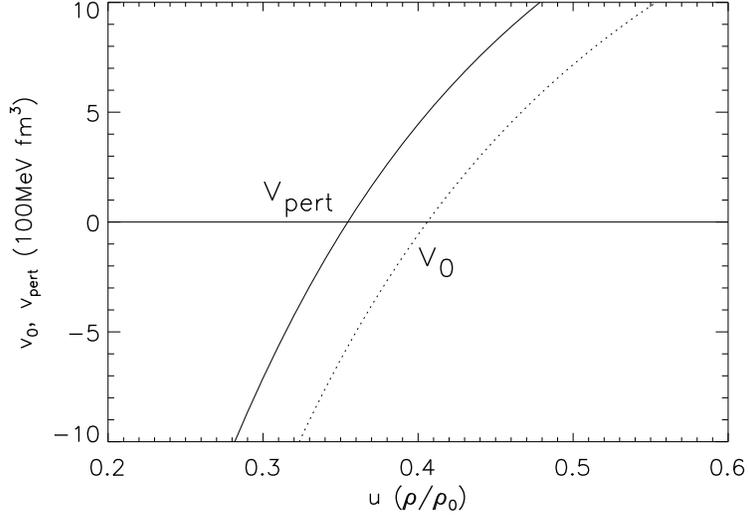}}
  \caption{We can see the transition density from nuclei with a neutron 
gas to nuclear matter. The solid line denotes the curve from the perturbation
 method. The dotted line corresponds to the thermodynamic instability method,
equilivalent to $v_{0}$ in the perturbation method. The solid line has a 
transition density of $0.355\rho_{0}$ and the dotted line has a transition 
density of $0.406\rho_{0}$.}
  \end{figure}
\end{center}

\subsection{Quark matter}
In this paper, we don't consider the appearance of hyperons since it is not
 clear how the hyperons and nucleons interact. Thus we simply consider
 the phase transition from uniform matter to quark matter. We use the MIT bag 
model for the quark matter equation of state. At T=0 MeV, the pressure and energy 
density are given by\cite{nobook}
\begin{equation}
\begin{aligned}
p & =-B + \sum_{f}\frac{1}{4\pi^{2}(\hbar c)^{3}}\biggl[\mu_{f}
(\mu_{f}^{2}-m_{f}^{2}c^{4})^{1/2}(\mu_{f}^{2}-\frac{5}{2}m_{f}^{2}c^{4}) 
+ \frac{3}{2}m_{f}^{4}c^{8}\ln \biggl(\frac{\mu_{f}
 +(\mu_{f}^{2}-m_{f}^{2}c^{4})^{1/2}}{m_{f}c^{2}}\biggr) \biggr]\\[8pt]
\epsilon & = B + \sum_{f}\frac{3}{4\pi^{2}(\hbar c)^{3}}\biggl[\mu_{f}
(\mu_{f}^{2}-m_{f}^{2}c^{4})^{1/2}(\mu_{f}^{2}-\frac{5}{2}m_{f}^{2}c^{4})
 - \frac{1}{2}m_{f}^{4}c^{8}\ln \biggl(\frac{\mu_{f} +(\mu_{f}^{2}
 -m_{f}^{2}c^{4})^{1/2}}{m_{f}c^{2}}\biggr) \biggr]\,,
\end{aligned}
\end{equation}
where the density for each quark flavor is given by
\begin{equation}
\rho_{f} =\frac{(\mu_{f}^{2} -m_{f}^{2})^{3/2}}{\pi^{2}(\hbar c)^{3}}\,.
\end{equation}
For the pure-quark phase we use $m_{u}=m_{d}=0$, $m_{s}=150 $MeV and 
$B=100$MeV fm$^{-3}$.
\begin{figure}\label{mixphase}
\begin{center}
\begin{tabular}{cc}
\scalebox{0.45}{\includegraphics{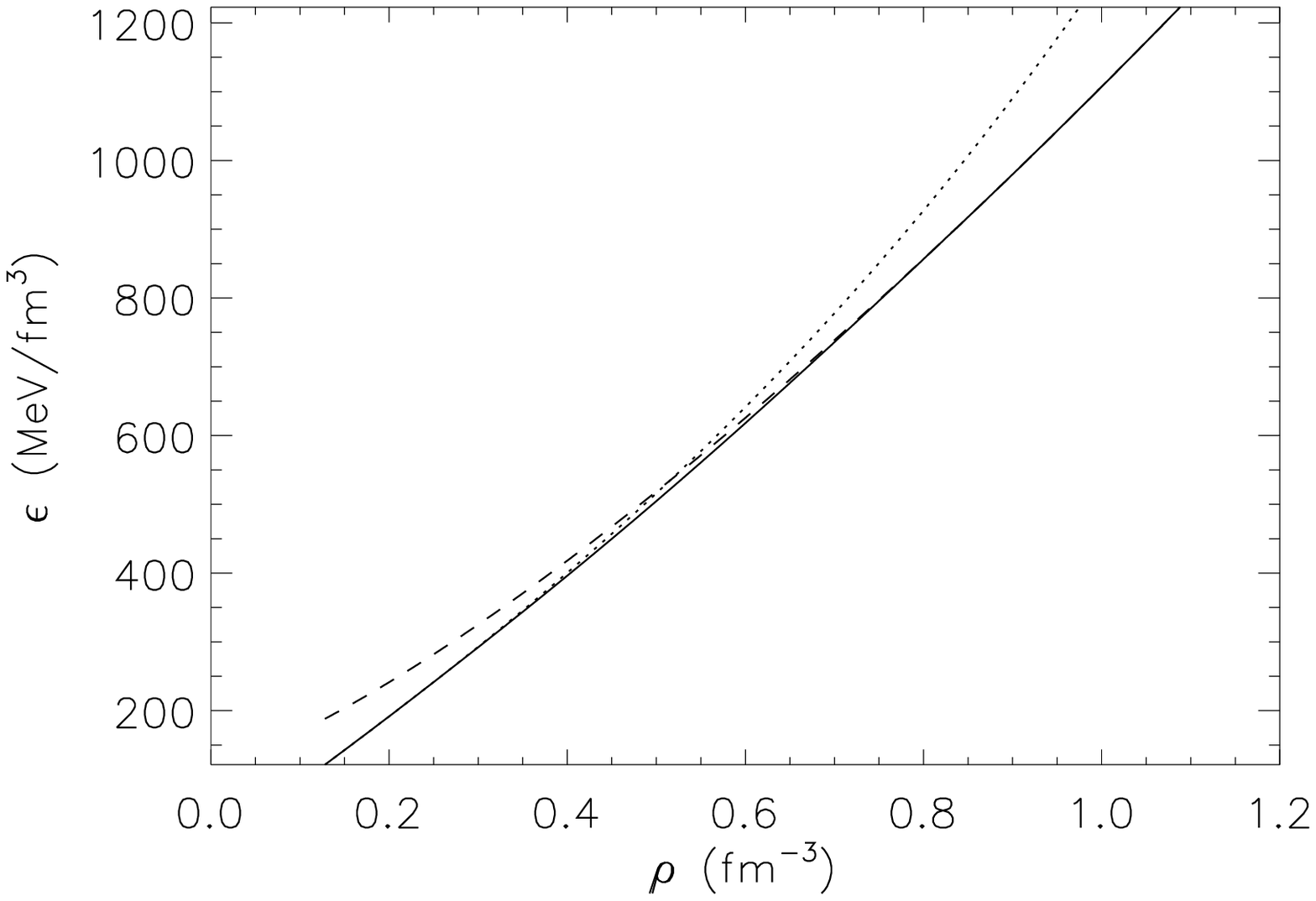}}&
\scalebox{0.45}{\includegraphics{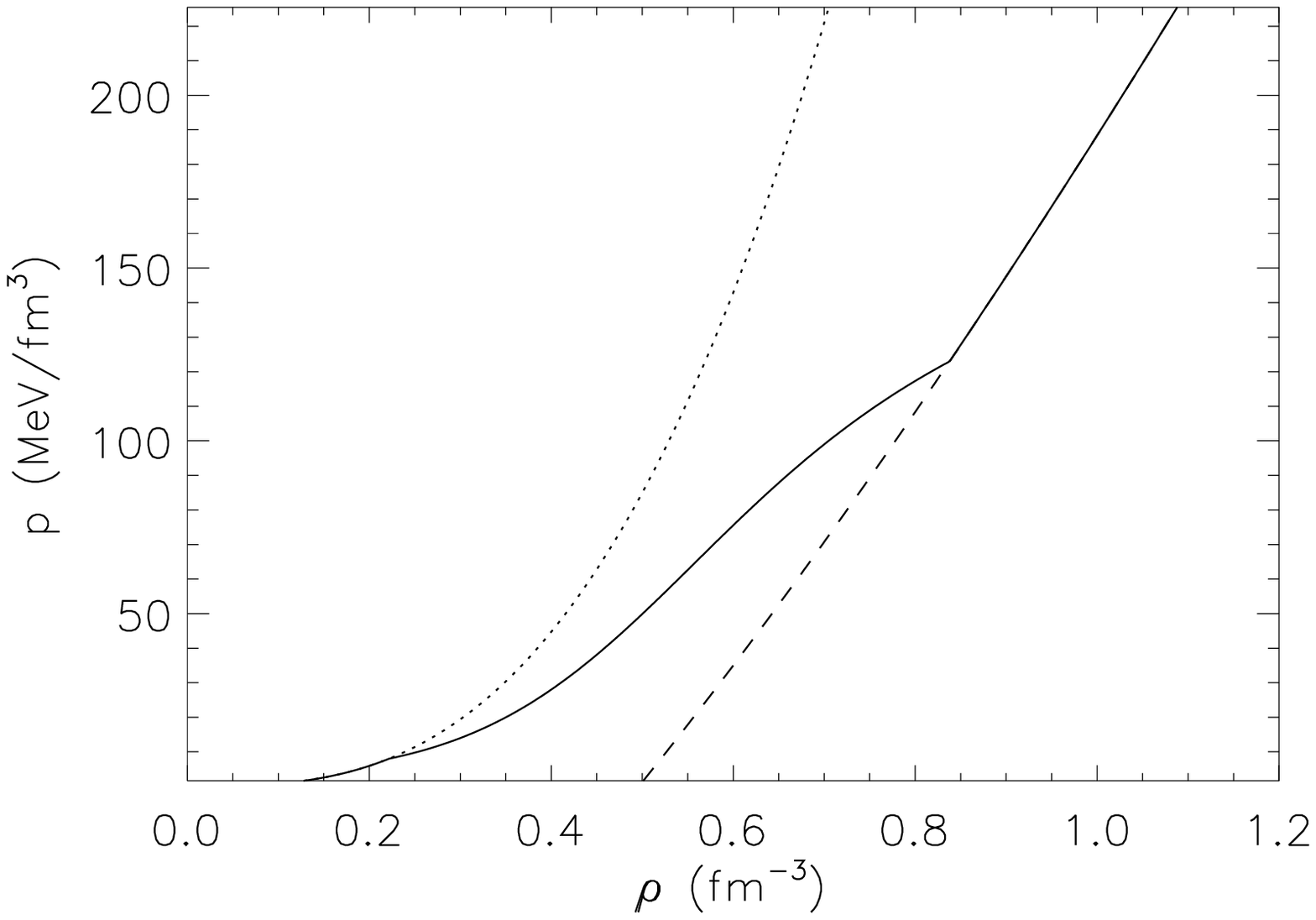}}
\end{tabular}
\caption{The left panel shows the energy density of nuclear matter
(dotted line), quark matter (dashed) and mixed phase (solid). The right
 panel shows the pressure of nuclear matter (dotted lined), 
quark matter (dashed) and mixed phase (solid).}
\end{center}
\end{figure}
In the mixed phase of uniform nuclear matter and quark matter, we apply
 Gibb's conditions to minimize the free-energy density with two constraints,
which are related to total number denstiy and charge neutrality.
\begin{equation}
\begin{aligned}
\rho_{b} & = \chi \rho_{N} + (1-\chi)\rho_{Q} \\
Q & =\chi Q_{N} + (1-\chi)Q_{Q}=0\,,
\end{aligned}
\end{equation}
where $\chi$ is the volume fraction of the uniform nuclear matter in the 
mixed phase and the subscript $N$ (Q) represents nuclear (quark) matter.
 In the mixed phase, the total charge is globally neutral in
contrast to pure nuclear matter and quark matter. From  minimizing  
the free enegy, we have
\begin{equation}
\begin{aligned}
p_{N} & = p_{Q} \\
\mu_{n} & = \mu_{u} + 2\mu_{d}\\
\mu_{p} & = 2\mu_{u} + \mu_{d}\,,\\
\end{aligned}
\end{equation}
then we have the energy density of the mixed phase
\begin{equation}
\epsilon  = \chi \epsilon_{N} + (1-\chi)\epsilon_{Q}\,. \\
\end{equation}
As in the case of uniform matter, we assume beta-stable matter so that
 the chemical potentials of the nuclear matter and quark matter have 
the relation
\begin{equation}
\begin{aligned}
\mu_{n} & =\mu_{p} + \mu_{e} \\
\mu_{d} & =\mu_{s}=\mu_{u}+\mu_{e}\,.
\end{aligned}
\end{equation}
Fig. 8 shows the energy density and pressure as a function of number 
density. The dotted (dashed) line represents nuclear (quark) matter. The
solid line denotes the mixed phase.The phase transition begins 
when the baryon density becomes 1.386$\rho_{0}$ 
and all nucleons turn into quark matter when the baryon density
 becomes 5.236$\rho_{0}$.\\
If there is a phase transition in the core of a cold neutron star,
 the mass and radius are quite different from the case of a 
pure-nuclear-matter  neutron star. 
The left panel of Fig. 9 shows the mass-radius relation of the neutron 
stars (dotted line) and hybrid stars (solid line). The right panel
shows the number density profiles of a quark matter and
 nuclear matter of 1.44$M_{\odot}$ hyrid star.
 The maximum mass of a cold neutron star with mixed 
phase (hybrid star) is $1.441 $M$_{\odot}$ and the central density of
 the neutron star is $9.585\rho_{0}$. The mass and radius curve with
 the mixed phase indicates when the mixed phase happens. 
\begin{figure}\label{sqmstar}
\begin{center}
\begin{tabular}{cc}
\scalebox{0.45}{\includegraphics{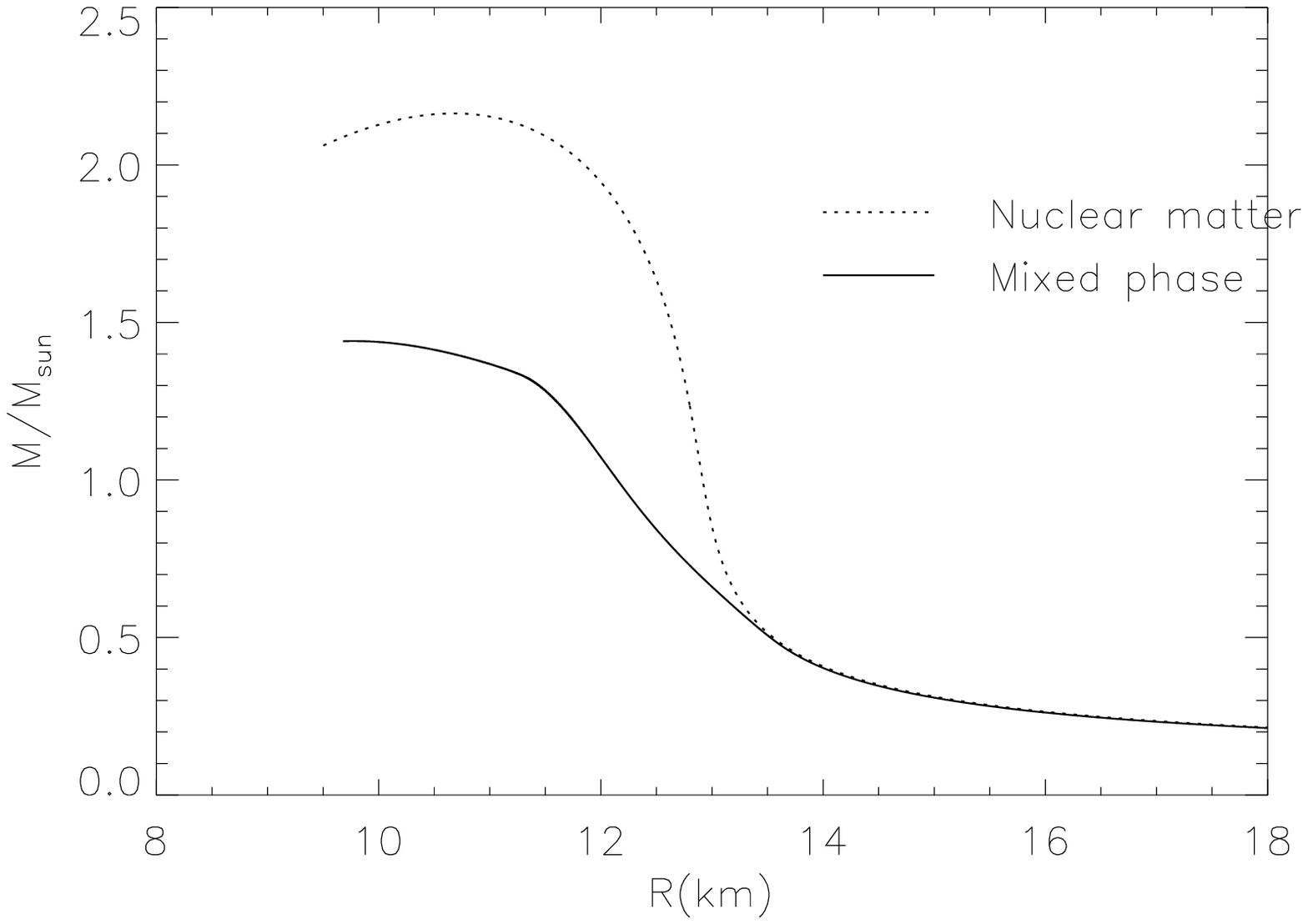}}&
\scalebox{0.45}{\includegraphics{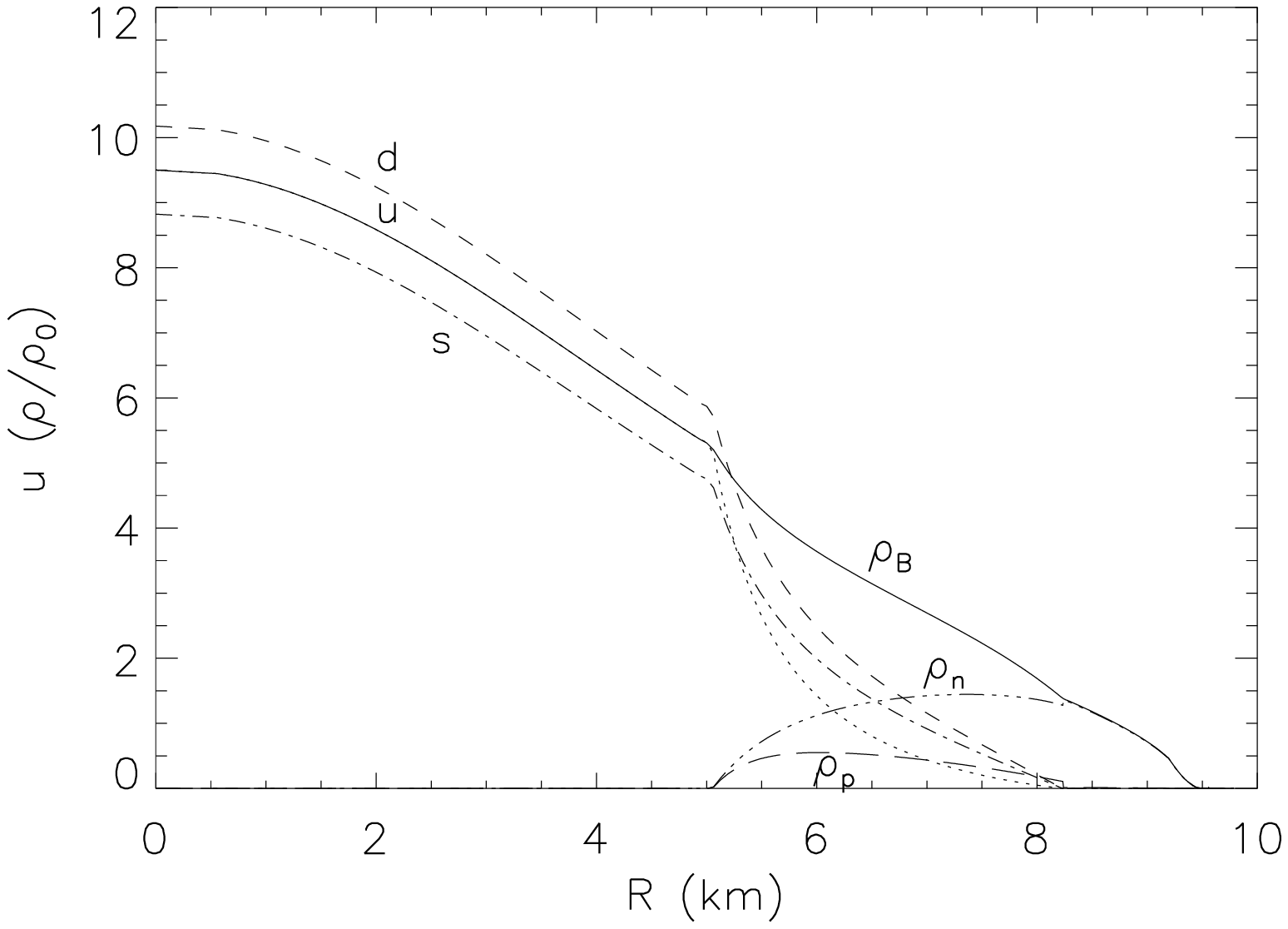}}
\end{tabular}
\caption{The left panel shows the mass-radius relation of the neutron 
stars (dotted line) from the GNDF and the hybrid stars (solid line). 
The right panel shows the 
number density profile of a hybrid star which have M= 1.44M$_{\odot}$.
At $r \sim$ 5.0km, the phase transition to nuclear matter takes place, 
and protons and neutrons exit. When r=$8.2$km, 
the phase transition to nuclear matter
is completed and there is no more quark matter. }
\end{center}
\end{figure}
As the distance from the center of the hybrid star increases, the phase
 transition to
nuclear matter takes place so that neutrons and protons appear($r \sim 5.0$km). 
Far from the center, the phase transition is completed ($r=8.2$km);
 pure nuclear matter exists only for larger radii.
The existence of quark matter in the hybrid star can be explained by 
angular momentum loss of the proto-neutron star. That is, fast-rotating
neutron stars lose angular momentum because of magnetic dipole radiation;
the central density of the neutron star increases due to the decrease in
centripetal forec, then quark matter appears.
 Since quark matter has a lower energy density than pure nuclear
 matter, we might expect heating of the neutron star from latent heat from quark 
 matter.

\section{Astrophysical application}
\subsection{Mass-radius relation of a cold neutron star}
We know that the radius of a neutron star is $\sim 10\text{km}$ and the 
mass is $\sim 1.4M_{\odot}$. In this system, the degeneracy pressure of the
 neutrons provides support against gravitational collapse. We can apply our
 model to calculate the mass and radius of neutron stars for a given central
 density. We use the Tolman-Oppenheimer-Volkov (TOV) equations which describe
 general relativistic hydrostatic equilibrium:
\begin{equation}\label{eq:tov}
\begin{aligned}
&\frac{dp}{dr} = - \frac{G(M(r)+4\pi r^{3} p/c^{2})(\epsilon + p)}
{r(r-2GM(r)/c^{2})c^{2}} \\[8pt]
&\frac{dM}{dr} = 4\pi \frac{\epsilon}{c^{2}}r^{2} \,.
\end{aligned}
\end{equation}
Fig. 10 shows the mass-radius relation for a cold neutron star. 
In the GNDF model, the maximum mass of a cold neutron star is 
$2.163$M$_{\odot}$, and the correspoding radius is $10.673$km. 
The maximum mass from the GNDF model is in between the FRTF truncated model
(FRTF I)
 and the modified model of the FRTF (FRTF II)\cite{frtf1}\cite{frtf2}.
\begin{center}
  \begin{figure}[h]
    \scalebox{0.6}{\includegraphics{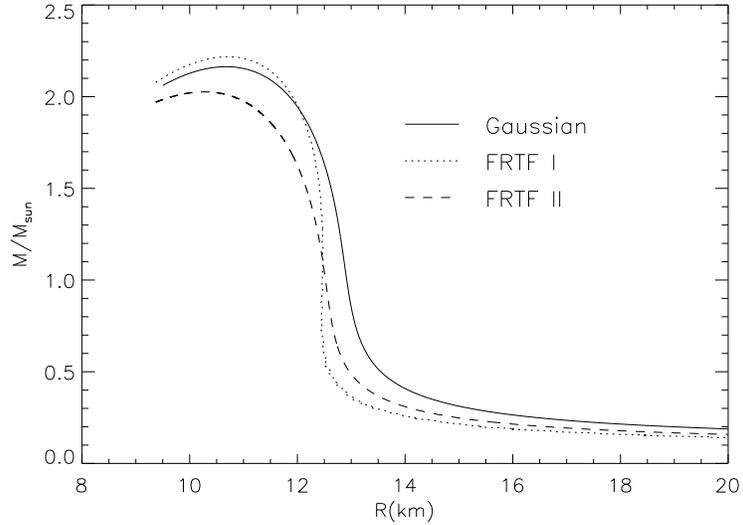}}
  \caption{Mass-radius relation for a cold neutron star.
 The mass of a cold neutron star from the
 Gaussian density functional model has a maximum mass of 
(2.163$M_{\odot}$) when the central density is 6.74$\rho_{0}$. }
  \end{figure}
\end{center}
\subsection{Moment of inertia of the neutron star}
In the slow-motion approximation, the moment of inertia is given 
by\cite{jl2000}
\begin{equation}
I=\frac{8\pi}{3}\int_{0}^{R} r^{4}(\rho+p)e^{(\lambda -\nu)/2}\omega \,dr\,,
\end{equation}
where $\lambda=-\ln(1-2m/r)$ and $\nu$ are the metric coefficients and $\omega$
 is the rotational drag function. In terms of the function 
$j=e^{-(\lambda +\nu)/2}$, the rotational drag statisfies
\begin{equation}\label{eq:omega2}
\frac{d}{dr}\biggl(r^{4}j\frac{d\omega}{dr}\biggr)=-4r^{3}\omega\frac{dj}{dr}\,,
\end{equation}
with the boundary conditions 
\begin{equation}
\omega_{R}=1-\frac{2I}{R^{3}}\,,\qquad \biggl(\frac{d\omega}{dr}\biggr)_{0}=0\,.
\end{equation}
Therefore, the moment of inertia can be written as
\begin{equation}
I=-\frac{2}{3}\int_{0}^{R}r^{3}\omega\frac{dj}{dr}dr=\frac{1}{6}\int_{0}^{R} 
d\biggl(r^{4}\omega \frac{dj}{dr}\biggr)=\frac{R^{4}}{6}\frac{d\omega}{dr}
\bigg{\vert}_{R}\,.
\end{equation}
We note that the second-order differential equation that $\omega$ satisfies,
 eq. (\ref{eq:omega2}), can be
 instead written as a first order differential equation in terms of the
 function $\phi =d\ln \omega /d \ln r$,
\begin{equation}
\frac{d\phi}{dr}=-\frac{\phi}{r}(\phi +3) -(4+\phi)\frac{d\ln j}{dr}\,,
\end{equation}
where 
\begin{equation}
\frac{d\ln j}{dr} =-\frac{4\pi r^{2}}{r-2m}(\rho + p)\,,
\end{equation}
with the boundary condition $\phi(0)=0$. The moment of inertia becomes
\begin{equation}\label{eq:emprical}
I=\frac{R^{3}}{6}\phi_{R}\omega_{R}=\frac{\phi_{R}}{6}(R^{3}-2I)\,,
\end{equation}
using the boundary condition for $\omega$. This simplifies to
\begin{equation}
I=\frac{R^{3}\phi_{R}}{6+2\phi_{R}}\,.
\end{equation}
Lattimer and Schutz proposed an empirical approximation for the moment of
 inertia\cite{ls2005},
\begin{equation}\label{eq:empirical}
I \simeq (0.237\pm0.008)MR^{2}\biggl[1 + 4.2\frac{M \text{km}}{\text{M}_{\odot}R}
 + 90\Big(\frac{M \text{km}}{\text{M}_{\odot}R}\Big)\biggr]\,.
\end{equation}
Fig. 11 shows the moment of inertia of a cold neutron star. The color band
represents upper and lower boundaries of the emprical approximation. 
FRTF I, II and GNDF agree quite well with this empirical approximation.

\begin{center}
  \begin{figure}[h]
    \scalebox{0.45}{\includegraphics[angle=-90]{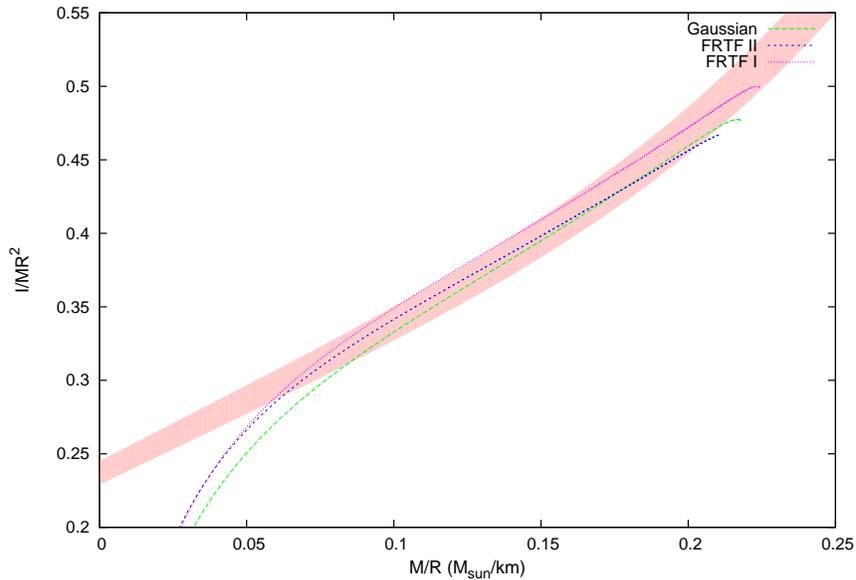}}
  \caption{Moment of inertia of a cold neutron star. The color band represents
the upper and lower boundaries of the emprical approximation 
eq (\ref{eq:empirical}). Three different models
  show different curves; however, they represent the emprical results quite 
  well.}
  \end{figure}
\end{center}

\subsection{Matter in the proto-neutron star}
We expect that the core of a proto-neutron star is lepton rich
 ($Y_{l} \sim 0.4$) and the entropy per baryon is $s \sim 2 - 4 $\cite{bl1986}.
 Since the Gaussian density functional model deals with non-relativistic
 nucleons, we use the relativistic leptonic formula to find the contribution 
from leptons to the total pressure, energy and entropy. 
Since the electrons and neutrinos are
 relativistic in proto-neutron star matter,
 we can get analytic solutions for leptons which are\cite{ls}
\begin{equation}
\mu_{l} = r - \frac{q}{r}\,, \quad r=[\sqrt{q^{3} + t^{2}} + t ]^{1/3}
\end{equation}
where $t=3\pi^{2} (\hbar c)^{3} n Y_{l}/g_{l}$ and 
$q=\frac{1}{3}(\pi T)^{2} -\frac{1}{2}m_{l}^{2}c^{4}$. Here $g_{l}$ is the spin
 degeneracy; thus $g_{e}=2$, and $g_{\nu}=1$ since the neutrino is only left
 handed. We assume that the neutrino is massless.\\
Expressions for the pressure and entropy per baryon are
\begin{equation}\label{eq:proto}
\begin{aligned}
p_{l} = & \frac{g_{l}\mu_{l}}{24 \pi^{2}}\left( \frac{\mu_{l}}{\hbar c}
\right)^{3}\Biggl[ 1 + \frac{2\pi^{2} T^{2} -3m_{l}^{2}c^{4}}{\mu_{l}^{2}}
+ \frac{\pi^{2}T^{2}}{\mu_{l}^{4}}\left(\frac{7}{15}\pi^{2}T^{2}
 -\frac{1}{2}m_{l}^{2}c^{4} \right) \Biggr] \,,\\[8pt]
s_{l} =& \frac{g_{l}T\mu_{l}^{2}}{6 n (\hbar c)^{3} }\left[ 1 
+  \frac{1}{\mu_{l}^{2}}\left(\frac{7}{15}\pi^{2}T^{2} 
-\frac{1}{2}m_{l}^{2}c^{4} \right) \right] \,.
\end{aligned}
\end{equation}
\begin{center}\label{protopre}
  \begin{figure}[h]
  \begin{tabular}{cc}
    \scalebox{0.45}{\includegraphics{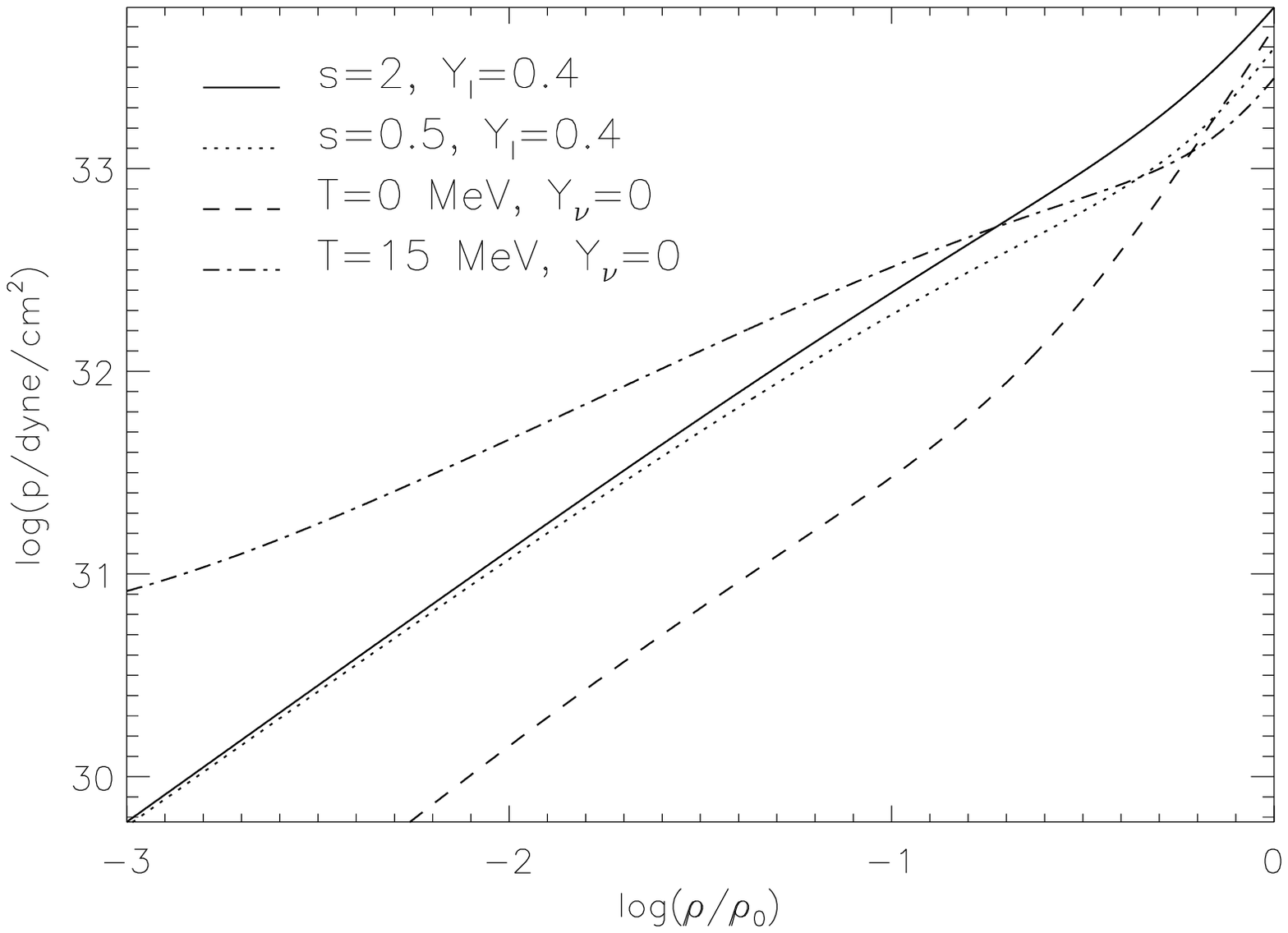}} &
    \scalebox{0.45}{\includegraphics{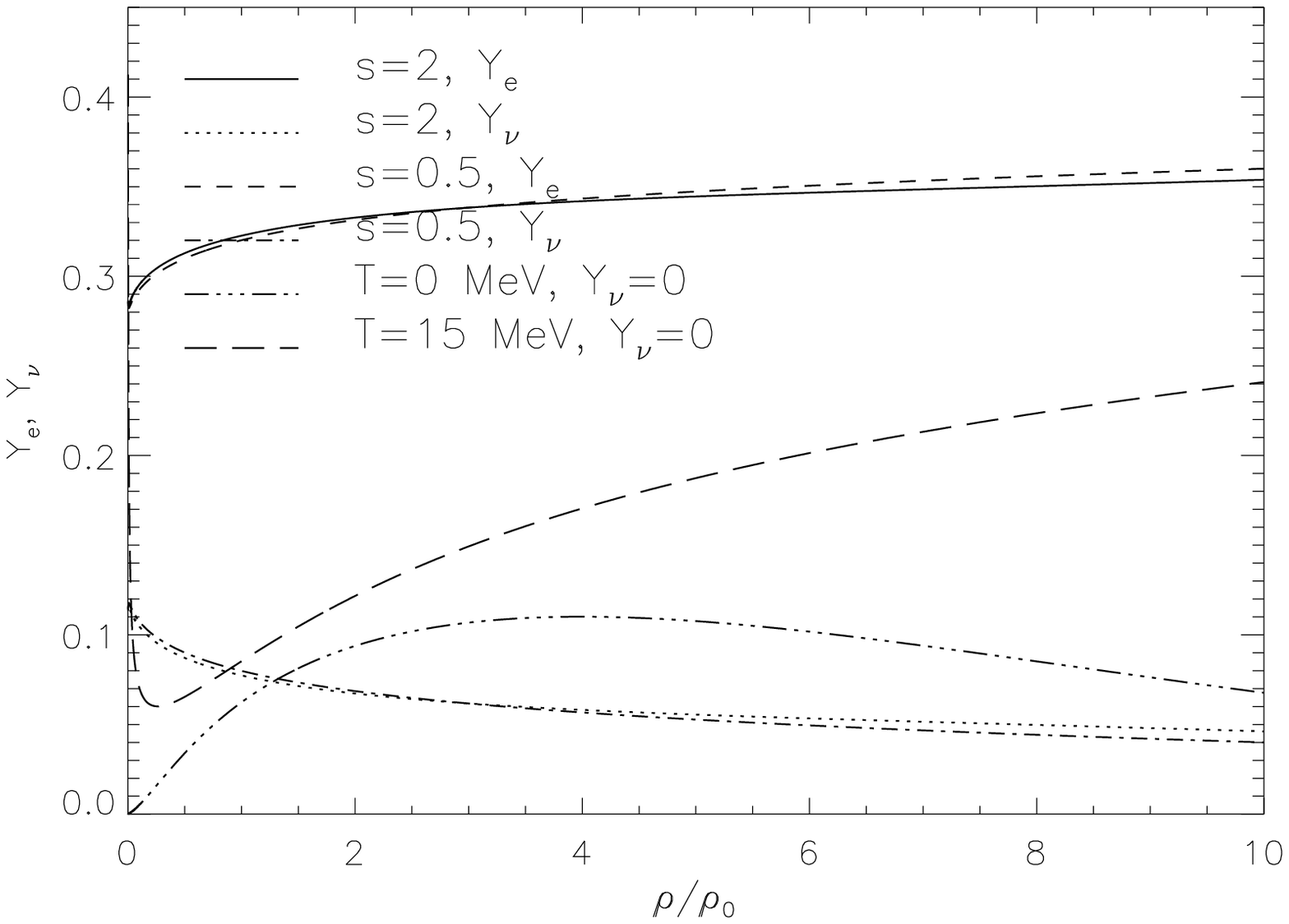}}
   \end{tabular}
  \caption{The left figure shows the total pressure when s=2.0, s=0.5, T=0MeV and 
   T=15MeV. The total pressure from various conditions shows different aspects.
    The total pressure for a fixed entropy per baryon (s=2, s=0.5) with fixed lepton 
   fraction (Y$_{l}$=0.4) shows identical results for lower densities, but they
   have different behavior as the density increases. The temperature for 
   a fixed entropy per baryon can be obtained from beta-equilibrium.
   The right figure shows the lepton fraction for four different cases. The results
   from fixed s=2 and s=4 are not distinguishable. The lepton fraction for T=0 
   and 15MeV, we assume beta-equilibrium.}
  \end{figure}
\end{center}
To simplify the model of the core of a proto-neutron star, 
we assume that there are only protons, neutrons,
 electrons and electron neutrinos . In this case
\begin{equation}\label{eq:proto2}
\begin{aligned}
s_{t} & =& s_{n} + s_{p} + s_{e} + s_{\nu_{e}} \\
Y_{l} & =& Y_{e} + Y_{\nu_{e}} \,.
\end{aligned}
\end{equation}
If the core of the proto neutron star is in equilibrium (to be
 more exact, it is in quasi-static equilibrium), then the EOS
 needs to meet eq. (\ref{eq:proto}) and eq. (\ref{eq:proto2}).
\begin{figure}\label{prots}
\begin{center}
\begin{tabular}{cc}
\scalebox{0.45}{\includegraphics{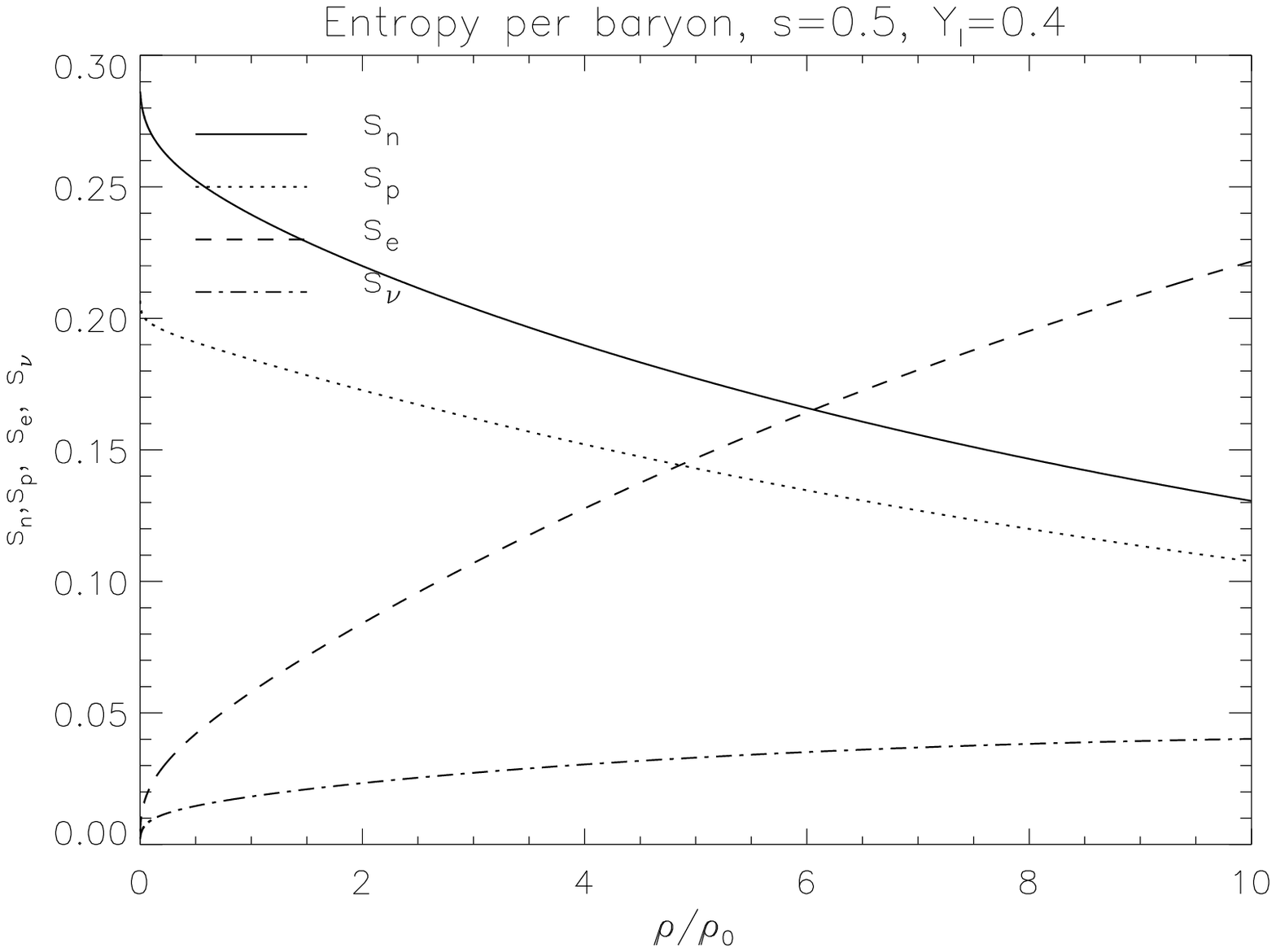}}&
\scalebox{0.45}{\includegraphics{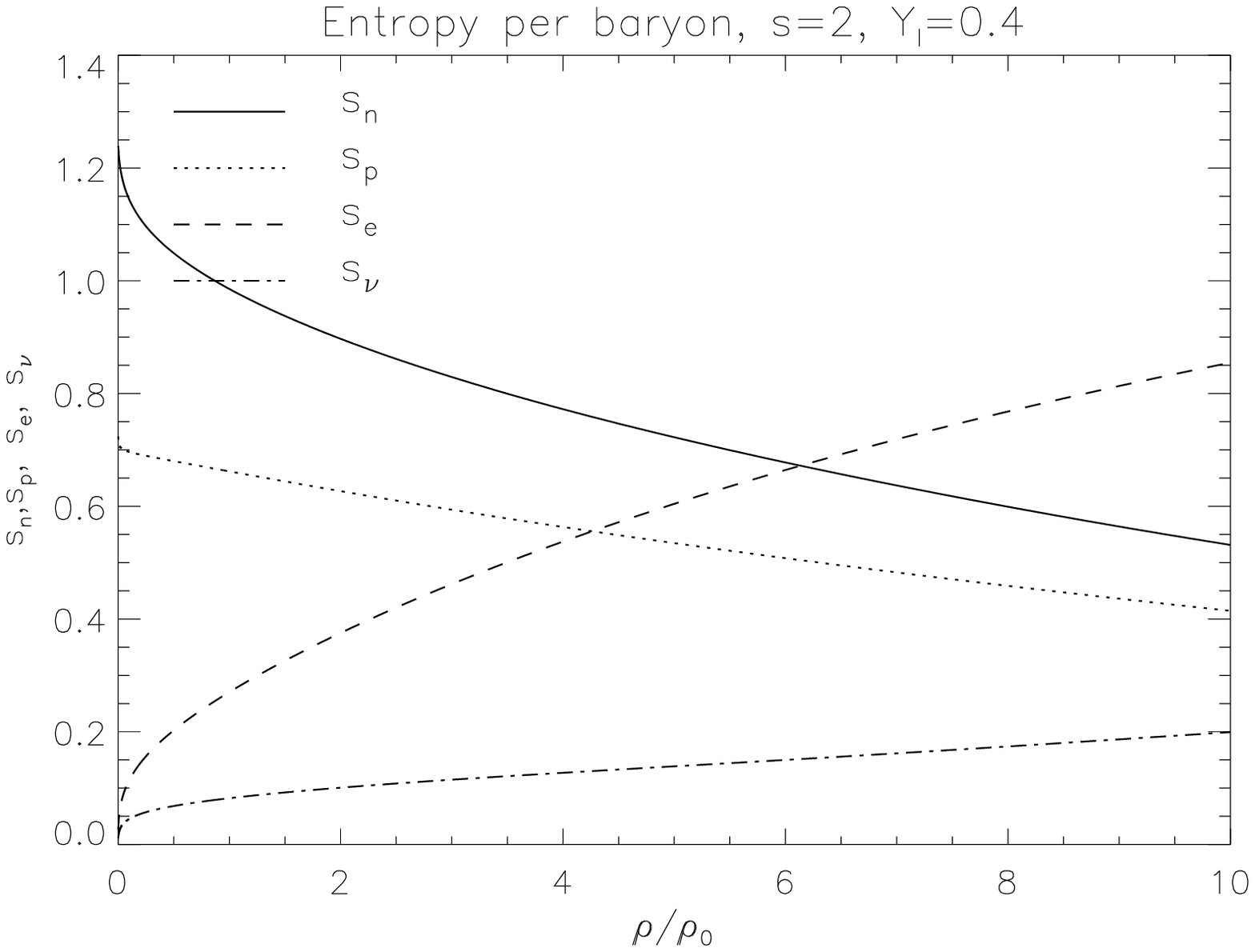}}
\end{tabular}
\caption{Entropy contribution from baryons and leptons for each fixed entropy
 per baryon. Since the neutron (electron) fraction is always greater than
 the proton (neutrino) fraction, the entropy contribution from neutrons 
(electrons) is greater than protons (neutrinos).}
\end{center}
\end{figure}
On the other hand, in the cold, catalysed neutron star matter, we can simply say
 that there are only electrons. Thus
\begin{equation}
\mu_{n} = \mu_{e} + \mu_{p} \,.
\end{equation}
If the interation of leptons with nuclear matter is weak, then the chemical 
potential and pressure of the leptons at zero temperature are given
 by\cite{ls},
\begin{equation}
\begin{aligned}
\mu_{l} =&\sqrt{(m_{l}c^{2})^{2} + (\hbar c)^{2}(3\pi^{2}\rho Y_{l})^{2/3}} \\
p_{l} =& \frac{1}{3\pi^{2}\hbar^{3}}\biggl[\sqrt{m^{2}c^{4} + p_{f}^{2}c^{2}}
\left(-\frac{3}{8}m^{2}p_{f}c^{5} + \frac{p_{f}^{3}c^{3}}{4} \right)
+ \frac{3}{8}m^{4}c^{8} \ln \Bigl(\frac{p_{f}c + \sqrt{m^{2}c^{4} + p_{f}^{2}c^
{2}}}{mc^{2}} \Bigr) \biggr]\,.
\end{aligned}
\end{equation}
In Fig. 12, the pressure is indistinguishable when s=2 and s=0.5 with
fixed lepton fraction (Y$_{l}=0.4$) at lower densities. However, 
for higher densities, they show a different behavior. 
At low densities, the pressure at high T is greater than the ones from the
fixed entropy per baryon with fixed lepton fraction, however, 
at high densities, the pressure contribution from the trapped neutrinos 
is much more important than the thermal effect. The right side of Fig. 12 shows
the lepton fraction from electrons and neutrinos for a given entropy per baryon.
The lepton fraction for each case is almost identical. For the cases T=0 and 15 MeV,
we assumed the beta-equilibrium condition. The electron fraction when T=15MeV 
increases from 0.06 to 0.25 between $\rho=0.3\rho_{0}$ $<$ $\rho=10\rho_{0}$.
Fig. 13 shows the various entropy contribution for a given entropy per baryon and
lepton fraction. Each graph shows the same trends. The entropy contributions from
baryons is decreasing as baryon density decreases, and the electron's (neutron's) 
contribution to entropy is always greater than the neutrino's (proton's)
 contribution 
since the electron (neutron) fraction is greater than the
neutrino (proton) fraction.\\

\section{conclusions}
In the GNDF model, the total energy consists of the
kinetic energy, the finite-range effect,
the zero-range effect, and the Coulomb energy. Using the
 Lagrange multiplier method, we find the potential energy, pressure, 
chemical potential and thermodynamic properties. The interaction parameters 
were obtained from the properties of infinite nuclear matter. 
We can find the charge radius and binding 
energy per baryon of the closed-shell nuclei. The Wigner-Seitz cell size 
increases as the density decreases and the binding energy per baryon approaches
-8.0MeV. The effective mass becomes 0.78$m$ at the center of the heavy nuclei and 
becomes $m$ outside of heavy nuclei. We were also able to find the pressure of 
uniform symmetric nuclear matter and neutron matter. For finite 
temperature, we can see the specific heat of nuclear matter follows the general 
trend of the free fermions. Thus the GNDF model is a good nuclear matter model 
to study for both low and high nuclear densities. We can improve the current model
 if we have more exact experimental results and we add additional interaction 
terms to explain the experimental results. The phase transition was studied  
using the GNDF model. The density perturbation suggested that the phase 
transition from non-uniform nuclear matter to uniform matter takes place
at densities less than $0.5\rho_{0}$. When we take into account the phase 
transition from uniform nuclear matter to quark matter, we see there is 
a drastic change in the maximum mass of a neutron star, since
 the pressure and energy density of quark matter are significantly different
from nuclear matter. The maximum mass 
of the hybrid star is less than $1.5$M$_{\odot}$. This might rule out the 
coexistence of nuclear matter and quark matter in compact stars.  The GNDF
model can  be used to study proto-neutron star matter
 combined with a leptonic environment. 
Using the GNDF theory,
we may compare the maximum and minimum masses of proto-neutron stars.\\
In a subsequent paper, we will study the nuclear 
pasta phase (spherical shell, cylinder, and slab geometry) using GNDF.\\

\section{Acknowledgements}
The author would like to thank James. M. Lattimer for useful discussion.
This work was supported in part by US DOE grant DE-FG02-87ER40317.

\end{document}